\documentclass[10pt]{article}
\usepackage{common}

\algdef{SE}[DOWHILE]{Do}{DoWhile}{\algorithmicdo}[1]{\algorithmicwhile\ #1}%

\title{\bf Self-Tuned Rejection Sampling within Gibbs and a Case Study in Small Area Estimation}

\author[a]{Andrew~M. Raim}
\author[a]{Kyle~M. Irimata}
\author[b]{James~A. Livsey}
\affil[a]{Center for Statistical Research and Methodology, U.S. Census Bureau}
\affil[b]{Department of Biostatistics, University at Buffalo}
\date{}

\renewcommand{\red}[1]{#1}

\begin{document}

\maketitle

\bigskip
\begin{abstract}
When formulating a Gibbs sampler, some conditionals may be unfamiliar distributions without well-known variate generation routines. Rejection sampling may be used to draw from such distributions exactly; however, it can be challenging to obtain practical proposal distributions. A practical proposal is one where accepted draws are not extremely rare occurrences and which is not too computationally intensive to use repeatedly within the Gibbs sampler. Consequently, approximate methods such as Metropolis-Hastings steps tend to be used in this setting. This work revisits the vertical weighted strips (VWS) method of proposal construction from \citet*{vws-2026} for univariate conditionals within Gibbs. VWS constructs a finite mixture based on the form of the target density and provides an upper bound on the rejection probability. The rejection probability can be reduced by refining terms in the finite mixture. Na\"{i}vely constructing a new proposal for each target encountered in a Gibbs sampler can be computationally impractical. Instead, we consider proposal distributions which persist over the Gibbs sampler and tune themselves gradually to avoid very high rejection probabilities while discarding mixture terms with low contribution. We explore a motivating application in small area estimation, applied to the estimation of county-level population counts of school-aged children in poverty. Here, a Gibbs sampler for a Bayesian model of interest includes a family of unfamiliar densities to be drawn for each observation in the data. Self-tuned VWS is applied to obtain exact draws within Gibbs while keeping the computational workload of proposal maintenance under control.
\end{abstract}

\noindent%

\blfootnote{%
Email: \texttt{andrew.raim@census.gov,
kyle.m.irimata@census.gov, 
jamesliv@buffalo.edu}. \\
Center for Statistical Research and Methodology,
U.S.~Census Bureau,
Washington, DC, 20233, U.S.A. \\
Department of Biostatistics,
University at Buffalo, Buffalo,
NY 14214, USA. \vspace{0.25em}}

\blfootnote{%
Disclaimer: This article is released to inform interested parties of ongoing research
and to encourage discussion of work in progress. Any views expressed are those of the
authors and not those of the U.S.~Census Bureau.}

\section{Introduction}
\label{sec:intro}

This work revisits the method of vertical weighted strips (VWS) for rejection sampling \citep*{vws-2026} and explores its use within a Gibbs sampler. VWS is a framework to construct proposals that regards the target distribution as a weighted density. The proposal takes the form of a finite mixture whose component densities correspond to a partition of the target distribution's support. This partition may be iteratively refined to reduce the rejection rate to a satisfactory threshold. This can often be used to achieve adequate-to-high acceptance rates when encountering unfamiliar targets. For univariate targets, which are the focus of \citet*{vws-2026}, an important multivariate application is to draw from conditionals in a Gibbs sampler which do not have well-known methods to directly generate variates. The refinement process can become computationally burdensome in this setting when repeated within many iterations of a Gibbs sampler, especially for latent variables corresponding to many subjects or groups. Here we consider a self-tuned variation of the VWS method. We utilize a rule of thumb to refine the proposal based on rejected draws and to coarsen it when we identify regions which do not substantially contribute to reducing the rejection rate. Intuitively, adjustments to the partition should become increasingly unnecessary as the chain from the Gibbs sampler approaches its invariant distribution.

As a motivating application, we focus on a Bayesian model for small area estimation (SAE). SAE is widely used in official statistics to augment estimates from sample surveys (``direct estimates'') with a model. In this context, ``areas'' are cross-sections of a population which may include both geographic and non-spatial characteristics such as age or race. ``Small areas'' are those cross-sections with small sample sizes where survey estimates may be unreliable. Consider a setting with $m$ areas, indexed $i = 1, \ldots, m$, each with direct point estimate $y_i$ and associated direct sampling variance estimate $s_i^2$ from $n_i$ survey respondents. SAE approaches based on the seminal work of \citet{Fay-Herriot-1979} typically utilize a mixed effects model to borrow strength among the areas through a common regression to provide more reliable estimates on small areas. The focus is often on modeling the point estimates $y_i$, while associated variance estimates $s_i^2$ are regarded as fixed; however, the $s_i^2$ are also estimates based on a survey sample and are subject to sampling variability. Inaccurate estimates of $s_i^2$ can lead to issues in modeling, especially for the mean squared error \citep{Bell-2008}. Two potential methods of accounting for uncertainty in variance estimates are sampling and smoothing methods \citep{You-2021}. Smoothing approaches utilize an external model to produce estimates of the sampling variances, which are in turn treated as known and used in a small area model. One such approach is the generalized variance function (GVF), which incorporates covariates to produce model-based estimates of the sampling variances. \citet{Otto-Bell-1995} propose the use of a GVF in a model for the sampling error covariance matrix in the U.S. Census Bureau's Current Population Survey. \citet{Hawala-2018} provide a comparison of different GVF approaches. \citet*{Maples-Bell-Huang-2009} consider the use of GVFs in the American Community Survey (ACS) and incorporate the corresponding estimates with an empirical Bayes approach. The degrees of freedom in modeling $s_i^2$ are given special consideration due to the complex survey design used in the ACS. 

Some authors take the approach to model point and variance direct estimates jointly. \citet*{Sugasawa-Tamae-Kubokawa-2017} propose a joint SAE model which includes both a regression and non-regression form for $s_i^2$ in a hierarchical Bayesian framework. These feature an explicit shrinkage formulation between direct and model-based variance estimates. \citet*{Parker-Holan-Janicki-2023} propose a variation of models in \citet{Sugasawa-Tamae-Kubokawa-2017} with a prior based on the multivariate log-gamma distribution; this yields a Gibbs sampler where all but one univariate conditional can be generated using standard distributions. \citet*{Savitsky-2023} introduce a Bayesian hierarchical model which uses a Poisson-lognormal formulation to jointly model population counts and the sampling variances. \citet{You-Hidiroglou-2023} consider joint SAE models in the context of estimating proportions. We will focus on a model presented by \citet{You-2021} based on a hierarchical Bayesian framework. This model includes a regression for sampling variances using an intercept and $n_i$; it appears straightforward to include additional covariates in this regression if desired. \citet{You-2021} proposes a Gibbs sampler where one family of conditionals has a nonstandard distribution; this will be the target for rejection sampling. \citet{You-2021} extends the work of \citet{You-Chapman-2006} which does not include a regression on $s_i^2$; a Gibbs sampler is obtained where all conditionals have standard distributions. \citet{You-2023} shows that an informative prior for the variance components in this hierarchical model is superior to a non-informative prior; thus, we likewise rely on an informative prior.

\citet{You-2021} uses independent Metropolis-Hastings (IMH) steps \citep[Section~7.4]{Robert-Casella-2004} to handle draws of the nonstandard conditionals. This is a standard algorithm choice, but we find that some of the associated chains in our application mix poorly. We instead consider self-tuned VWS to obtain exact draws of these conditionals which is seen to substantially improve slower mixing chains. Self-tuned VWS requires more computation and bookkeeping than IMH and other variants of a Metropolis step, but performs well enough to be practical in a setting such as our motivating application.

Properly constructed MCMC chains are known to converge to their intended target, but their progress in a finite number of iterations is not as well understood. Diagnostic techniques are often used to assess convergence in practice \citep{Cowles-Carlin-1996, Roy-2020}. Two major motivations for such diagnostics are mentioned by \citet*{BDA3}: draws from a chain may be unrepresentative of the target when far from convergence, and may be highly correlated due to slow mixing. In both cases, their use to approximate aspects of the target distribution may be misleading. A number of factors in the design of a Gibbs sampler can potentially influence its mixing. \citet{Roberts-Sahu-1997} study updating strategies including use of deterministic sequences, random selection of a conditional to update, and blocking. \citet{vanDyk-Park-2008} consider marginalizing the joint distribution of a target posterior or some of its conditionals via partial collapsing. \citet{Liu-1994} and \citet{vanDyk-Meng-2001} consider the effect of data augmentation on mixing. \citet*{Ascolani-Roberts-Giacomo-2026} and \citet*{Qin-Ju-Wang-2025} develop theory to compare mixing under exact generation from conditionals versus use of Metropolis steps. One reason IMH may be preferred to random walk Metropolis-Hastings is that properties of the former are less challenging to analyze---proposals from the latter depend on the position of the chain---though the former may not mix as well \citep{Liu-1996, Wang-2022}.

Within a Gibbs sampler, rejection sampling can be used to draw exactly from a conditional which does not have a recognizable form. This avoids issues of convergence and autocorrelation encountered when using Metropolis steps; however, it can be challenging to formulate a proposal that accepts with high enough probability and also avoids an impractical amount of computation when repeated in the Gibbs sampler. Griddy Gibbs \citep{Ritter-Tanner-1992} forms an approximation to the CDF of a target that can be sampled with the inverse CDF method. The grid can persist throughout Gibbs sampling and be tuned periodically. Here, the invariant distribution of the Gibbs sampler depends on the accuracy of approximations to conditionals. Within-Gibbs sampling was a major consideration in the development of the adaptive rejection sampling (ARS) and adaptive rejection Metropolis sampling (ARMS) methods \citep*{Gilks-Wild-1992, Gilks-Best-Tan-1995}. To reduce effort spent in proposal construction, these works make use of quantiles from the proposal in the previous Gibbs iteration as initial abscissae in the next iteration. The Fast Universal Self-tuned Sampler (FUSS) proposed by \citet*{Martino-EtAl-2015} lifts the ARS restriction of log-concavity for the target. FUSS starts with a large pool of candidate support points and selects a subset during initialization before proceeding with sampling. FUSS ensures that the original target is the invariant distribution of the overall Gibbs sampler. One variant uses a rejection sampling step followed by a Metropolis step, similar to ARMS, to avoid large correlation among the conditionals. Cheap Adaptive Rejection Sampling (CARS) proposed by \citet{Martino-Louzada-2019} controls the amount of computation in ARS by assuming a fixed number of support points which can be modified throughout sampling. \citet{James-2024} revisits ARS and proposes choosing a small number of initial support points surrounding the mode of a log-concave density. The points can be quickly computed to avoid overhead when repeated within a Gibbs sampler. The present work seeks to obtain exact samples from unfamiliar conditionals which are not necessarily log-concave using rejection sampling within Gibbs. Proposals are automatically tuned as the chain evolves to keep rejection probability at a threshold below 1 but avoid incurring a large computational overhead from many such adjustments.

The remainder of the paper proceeds as follows. Section~\ref{sec:saipe-data} presents additional details about the SAE setting and introduces a dataset which will serve as a motivating example. Section~\ref{sec:you} introduces the joint SAE model and the IMH-within-Gibbs method originally proposed to fit it. Section~\ref{sec:self-tune-vws} presents the self-tuned VWS method. Section~\ref{sec:vws-within-gibbs} derives expressions to implement self-tuned VWS-within-Gibbs for the joint SAE application. Section~\ref{sec:sim} presents simulation studies comparing the samplers. Section~\ref{sec:saipe-analysis} presents an analysis of the example data from Section~\ref{sec:saipe-data}. Section~\ref{sec:conclusions} discusses findings and concludes the paper.

An electronic supplement and replication materials are provided in \citet*{zenodo-2026}. The supplement includes additional results for Sections~\ref{sec:sim} and \ref{sec:saipe-analysis}, as well as a second Bayesian application in SAE illustrating self-tuned VWS in a Gibbs sampler. Table and figure references in the present article with an ``S'' prefix refer to the supplement. Sampling methods in the replication materials are implemented in C++ and invoked from R via Rcpp \citep{Eddelbuettel-2013}. Support for self-tuned VWS is available in the \texttt{vws} R package \citep{vws}, as of version 0.4.0, which is utilized in the replication codes. Run times reported in the two data analysis applications---Sections~\ref{sec:saipe-analysis} and S5---and in the conditional simulation study were measured on an AMD Ryzen 5 5600G workstation with six CPU cores and 16~MB of L3 cache. The posterior simulation study in Section~\ref{sec:sim-post} was run on a single node of a computing cluster with two Intel Xeon Gold 6142 CPUs, each with 16 cores and 22~MB of L3 cache. 

\section{SAIPE Dataset}
\label{sec:saipe-data}

One widely known implementation of SAE is the Small Area Income and Poverty Estimates (SAIPE) program administered by the U.S. Census Bureau. This program produces annual estimates of income and poverty for states, counties, and school districts in the U.S. \citep*{Bell-EtAl-2016}. One measurement of interest is the county-level count of the population in poverty for school-aged children (age 5--17). The ACS provides primary estimates of poverty for SAIPE via direct point estimates and sampling variances which are based on the survey design. SAIPE incorporates auxiliary data sources into the production model as covariates, including federal income tax data from the U.S. Internal Revenue Service, Supplemental Nutrition Assistance Program (SNAP) data, and population counts from the U.S. Census Bureau's Population Estimates Program (PEP). \citet{Bell-EtAl-2016} provide a comprehensive review of the SAIPE methodology and covariates used for the production model.

In practice, SAIPE utilizes 1-year ACS data to produce poverty estimates; however, this granularity of ACS data is not publicly released for counties with populations under 65,000. For this work, we utilize an analogous data set which is publicly available. Poverty estimates for counties in the U.S. were obtained from the 2023 5-year ACS table S1701: ``Poverty Status in the Past 1 Months''; this includes point estimates and associated sampling variance estimates based on published margins of error. Denote the point and variance estimates on their originally published scale as $\tilde{y}_i$ and $\tilde{s}_i^2$, respectively, for counties $i = 1, \ldots, m$. Counts of sampled housing units $n_i$ were obtained from the 5-year 2023 ACS table B98001: ``Unweighted Housing Unit Sample''. Because the distribution of estimates $\tilde{y}_1, \ldots, \tilde{y}_m$ is right-skewed, they are typically transformed to $y_i = \log \tilde{y}_i$ for modeling \citep{Bell-EtAl-2016}. Mimicking the SAIPE production model used by the U.S. Census Bureau, we obtain county level predictors for the count of participants in SNAP in 2022, as well as population estimates for 2023 from the 2020--2023 vintage of PEP. We denote the corresponding predictors for county $i$ by $\text{SNAP}_i$ and $\text{PEP}_i$, respectively. Other covariate data sources used in SAIPE---such as federal income tax data---are not publicly available and omitted from our dataset. We note that this tax data is one of the primary data sources for the production SAIPE model, thus we expect that the results from this example will not be directly comparable to the official SAIPE release.

The variance estimates $\tilde{s}_i^2$ must be altered to correspond to the log-transformed point estimates. Statisticians within the U.S. Census Bureau can accomplish this for the official releases by applying a set of replicate weights to the transformed outcomes \citep{Huang-Bell-2009}. For this work, we instead use the delta method to make use only of publicly released data. Suppose $\tilde{y}_i$ has unknown mean $\tilde{\vartheta}_i$ and variance $\tilde{\sigma}_i^2$ and consider the first-order Taylor expansion
\begin{math}
\log \tilde{y}_i
\approx \log \tilde{\vartheta}_i + \tilde{\vartheta}_i^{-1} (\tilde{y}_i - \tilde{\vartheta}_i)
\end{math}
with approximate variance
\begin{math}
\Var( \log \tilde{y}_i) \approx \tilde{\sigma}_i^2 / \tilde{\vartheta}_i^2.
\end{math}
Plugging in the observed $\tilde{y}_i$ for $\tilde{\vartheta}_i$ and
$\tilde{s}_i^2$ for $\tilde{\sigma}_i^2$ yields a transformed variance estimate
\begin{math}
s_i^2 = \tilde{s}_i^2 / \tilde{y}_i^2.
\end{math}
The quantities of primary interest in modeling will be the underlying mean $\vartheta_i$ and variance $\sigma_i^2$ of the transformed $y_i$.

\begin{figure}
\centering
\caption{Summary of $\sigma_i^2$ draws from posterior of joint small area estimation model fitted in Section~\ref{sec:saipe-analysis} using VWS. (\subref{fig:modeled-vs-sigma2}) Point estimates of $\hat{\sigma}_i^2$ versus $s_i^2$. (\subref{fig:variance-ci-width}) Width of 90\% credible interval for $\sigma_i^2$ versus $\log n_i$.}
\label{fig:joint-sigma2}
\begin{subfigure}[t]{0.35\textwidth}
\centering
\includegraphics[width=\textwidth]{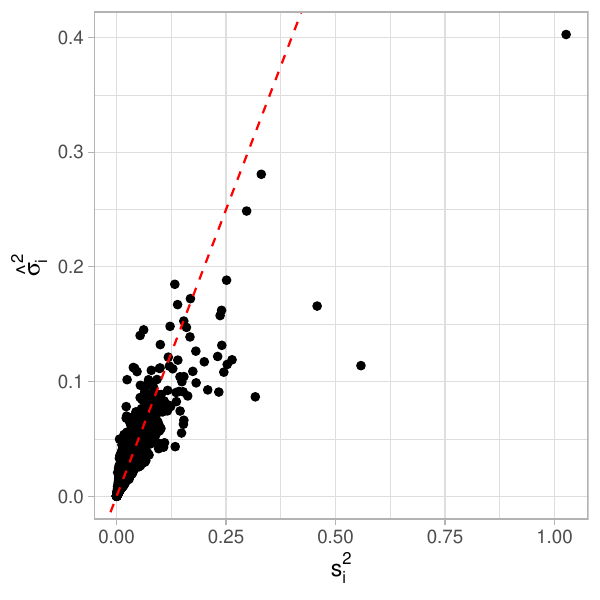}
\caption{}
\label{fig:modeled-vs-sigma2}
\end{subfigure}
\begin{subfigure}[t]{0.35\textwidth}
\centering
\includegraphics[width=\textwidth]{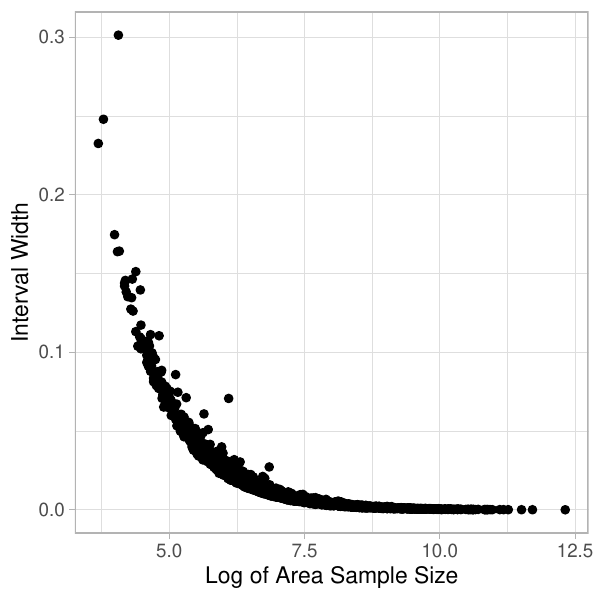}
\caption{}
\label{fig:variance-ci-width}
\end{subfigure}
\end{figure}

\begin{figure}
\centering
\caption{Ratio for 90\% credible interval widths for $\vartheta_i$ between Fay-Herriot and joint models. Horizontal axis displays the logarithm of area-level sample size $n_i$.}
\label{fig:theta-ci-ratio}
\includegraphics[width=0.50\textwidth]{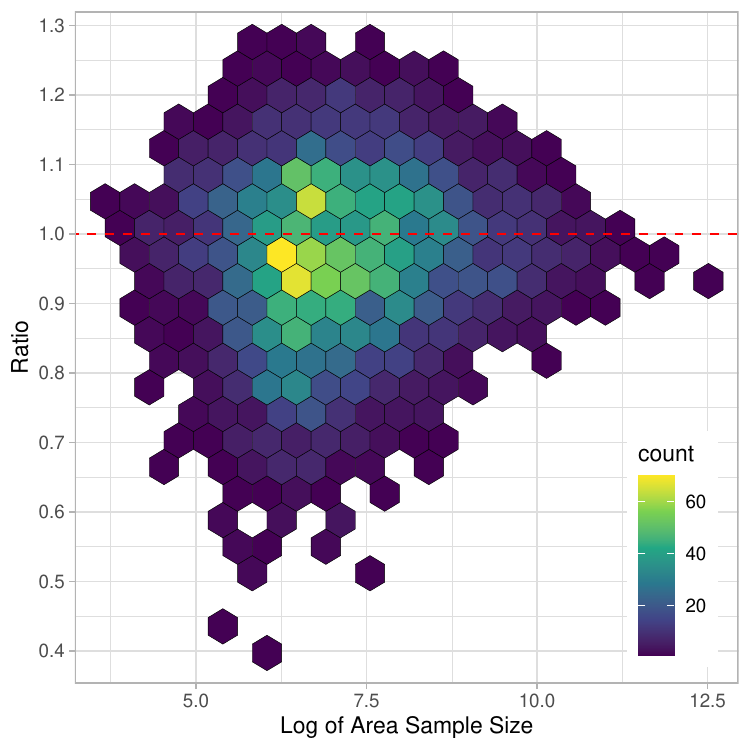}
\end{figure}

Let us consider several plots to briefly motivate the recognition of uncertainty in the sampling variances. Figure~\ref{fig:joint-sigma2} displays point estimates and interval widths from the posterior distribution of $\sigma_i^2$ under the joint model of \citet{You-2021}. Comparing point estimates $\hat{\sigma}_i^2$ from the model to direct variance estimates $s_i^2$ in Figure~\ref{fig:modeled-vs-sigma2} shows that they may be somewhat different. While the $\hat{\sigma}_i^2$ are sometimes larger than the $s_i^2$, the largest $s_i^2$, such as those greater than 0.25, are shrunken to smaller $\hat{\sigma}_i^2$. Figure~\ref{fig:variance-ci-width} shows that widths of 90\% credible intervals for $\sigma_i^2$ decrease rapidly with the logarithm of $n_i$. Larger uncertainty is expressed in areas with smaller $n_i$ as we may have anticipated; here it is more consequential to ignore this variability by regarding the $s_i^2$ as fixed. Focusing on $\vartheta_i$, point estimates (not shown) are similar between the joint model and the standard Fay-Herriot model \citep{Fay-Herriot-1979}. However, expressed uncertainties are more substantial; Figure~\ref{fig:theta-ci-ratio} compares widths of 90\% credible intervals for $\vartheta_i$ from the two models. Intervals are computed using 0.05 and 0.95 sample quantiles of draws taken from the posterior. A ratio of widths for each $\vartheta_i$ is formed using the Fay-Herriot model in the numerator and joint model in the denominator; therefore, values smaller than 1 indicate higher variability expressed by the joint model. The ratios are plotted against $\log n_i$ to capture their relationship to the sample size. We find that the proportion of ratios above and below 1 are generally similar; however, we also see that the joint model has less variability in the most extreme cases compared to the Fay-Herriot model. Ratios deviating farther from 1 appear most often in counties with moderate sample sizes.

\section{A Joint SAE Model and Gibbs Sampler}
\label{sec:you}

Model~3 of \citet{You-2021} jointly models direct point estimates and associated sampling variances using a likelihood based on
\begin{align}
y_i \sim \text{N}(\vartheta_i, \sigma_i^2), \quad
\vartheta_i \sim \text{N}(\vec{x}_i^\top \vec{\beta}, \phi^2), \quad
d_i s_i^2 / \sigma_i^2 \sim \chi_{d_i}^2, \quad
\log \sigma_i^2 \sim \text{N}(\vec{z}_i^\top \vec{\gamma}, \tau^2),
\label{eqn:model}
\end{align}
with a regression on both the latent mean $\vartheta_i$ and latent variance $\sigma_i^2$ of $y_i$. The degrees of freedom $d_i$ would be $n_i - 1$ under simple random sampling but require additional consideration under a complex survey design such as in the ACS (see Section~\ref{sec:saipe-analysis}). The model specification is completed by assuming flat priors with
$\pi(\vec{\beta}) \propto 1$,
$\pi(\vec{\gamma}) \propto 1$,
$\pi(\phi^2) \propto 1$, and
$\pi(\tau^2) \propto 1$.
Denote $\vec{y} = (y_1, \ldots, y_m)^\top$, $\vec{s}^2 = (s_1^2, \ldots, s_m^2)^\top$, $\vec{\vartheta} = (\vartheta_1, \ldots, \vartheta_m)^\top$, $\vec{\sigma}^2 = (\sigma_1^2, \ldots \sigma_m^2)^\top$, $\vec{X} = (\vec{x}_1 \, \cdots \, \vec{x}_n)^\top$, and $\vec{Z} = (\vec{z}_1 \, \cdots \, \vec{z}_n)^\top$. Here, $\vec{y}$, $\vec{s}^2$, $\vec{X}$, $\vec{Z}$, and $d_1, \ldots, d_m$ are observed, $\vec{\vartheta}$ and $\vec{\sigma}^2$ are latent random variables, and $\vec{\theta} = (\vec{\beta}, \vec{\gamma}, \phi^2, \tau^2)^\top$ are model parameters. 
Let $f_\text{N}(\vec{x} \mid \vec{\mu}, \vec{\Sigma})$ be the normal density,
$f_\text{LN}(x \mid \mu, \sigma^2)$ be a lognormal density with location $\mu$ and scale $\sigma$,
$f_\text{Ga}(x \mid a, b) \propto x^{a-1} e^{-bx}$ be the Gamma density with shape parameter $a$ and rate parameter $b$, and
$f_\text{IG}(x \mid a, b) \propto x^{-a-1} e^{-b/x}$ be the Inverse-Gamma density with shape parameter $a$ and rate parameter $b$.

The Gibbs sampler presented by \citet{You-2021} is obtained by regarding $\vec{\vartheta}$ and $\vec{\sigma}^2$ as augmented data. The joint distribution of all random variables is
\begin{align}
\pi(\vec{y}, \vec{s}^2, \vec{\vartheta}, \vec{\sigma}^2, \vec{\beta}, \vec{\gamma},
\phi^2, \tau^2)
&= \prod_{i=1}^m f_\text{N}(y_i \mid \vartheta_i, \sigma_i^2) \cdot
\prod_{i=1}^m f_\text{Ga}(d_i s_i^2 / \sigma_i^2 \mid d_i / 2, 1 / 2) \nonumber \\
&\times f_\text{N}(\vec{\vartheta} \mid \vec{X} \vec{\beta}, \phi^2 \vec{I}) \cdot
f_\text{N}(\log \vec{\sigma}^2 \mid \vec{Z} \vec{\gamma}, \tau^2 \vec{I}) \cdot
\prod_{i=1}^m \frac{1}{\sigma_i^2},
\label{eqn:full-joint}
\end{align}
where $\log \vec{\sigma}^2 = (\log \sigma_1^2, \ldots, \log \sigma_m^2)^\top$. Conditionals of \eqref{eqn:full-joint} may be used to form a Gibbs sampler for the data-augmented posterior
\begin{math}
\pi(\vec{\vartheta}, \vec{\sigma}^2, \vec{\beta}, \vec{\gamma}, \phi^2, \tau^2 \mid \vec{y}, \vec{s}^2);
\end{math}
these conditionals are given in Algorithm~\ref{alg:you}. In particular, we scan sequentially through steps 1--6 and repeat for a given number of iterations $R$. With the exception of step~\ref{step:target}, generating from the remaining conditionals is straightforward; therefore, let us focus on step~\ref{step:target}.

\begin{algorithm}
\caption{Conditionals for a Gibbs sampler based on joint SAE model.}
\label{alg:you}
\begin{enumerate}
\item $[\vec{\vartheta} \mid \rest]$ is a product of independent univariate conditionals with $[\vartheta_i \mid \rest] \sim \text{N}(p_i y_i + (1 - p_i) \vec{x}_i^\top \vec{\beta}, p_i \sigma_i^2)$ and $p_i = \phi^2 / (\phi^2 + \sigma_i^2)$.

\item $[\vec{\beta} \mid \rest] \sim \text{N}\left((\vec{X}^\top \vec{X})^{-1} \vec{X}^\top \vec{\vartheta}, \phi^2 (\vec{X}^\top \vec{X})^{-1}\right)$.

\item $[\vec{\gamma} \mid \rest] \sim \text{N}\left((\vec{Z}^\top \vec{Z})^{-1} \vec{Z}^\top \{\log \vec{\sigma}^2\}, \tau^2 (\vec{Z}^\top \vec{Z})^{-1}\right)$.

\item $[\phi^2 \mid \rest] \sim \text{IG}(m/2 - 1, \frac{1}{2} \lVert \vec{\vartheta} - \vec{X} \vec{\beta} \rVert^2)$.

\item $[\tau^2 \mid \rest] \sim \text{IG}(m/2 - 1, \frac{1}{2} \lVert \log \vec{\sigma}^2 - \vec{Z} \vec{\gamma} \rVert^2)$.

\item \label{step:target}
\begin{math}
[\vec{\sigma}^2 \mid \rest]
\end{math}
is a product of independent univariate conditionals with densities
\begin{align}
f(\sigma_i^2 \mid \rest) &\propto
f_\text{N}(y_i \mid \vartheta_i, \sigma_i^2)
f_\text{LN}(\sigma_i^2 \mid \vec{z}_i^\top \vec{\gamma}, \tau^2)
f_{\text{Ga}}(d_i s_i^2 / \sigma_i^2 \mid d_i/2, 1/2)
\nonumber \\
&\propto 
f_\text{IG}(\sigma_i^2 \mid \kappa_i, \lambda_i)
f_\text{LN}(\sigma_i^2 \mid \mu_i, \tau^2)
\label{eqn:target}
\end{align}
where $\mu_i = \vec{z}_i^\top \vec{\gamma}$, $\kappa_i = (d_i-1)/2$ and $\lambda_i = (y_i - \vartheta_i)^2 / 2 + d_i s_i^2 / 2$.
\end{enumerate}
\end{algorithm}

\begin{remark}
\label{remark:concavity}
Density \eqref{eqn:target} is not log-concave on $[0, \infty)$; therefore, the ARS algorithm and other sampling methods that rely on this property cannot be used directly. To see this, let
\begin{math}
\zeta(x) = \log f(x) = -x - \log x -\frac{1}{2 \tau^2} (\log x - \vartheta)^2.
\end{math}
We have that
\begin{math}
\zeta'(\omega^2) = -1 - x^{-1} \tau^{-2} (\log x - \vartheta + \tau^2),
\end{math}
and
\begin{math}
\zeta''(x) = x^{-2} \tau^{-2} (\log x - \vartheta - 1 - \tau^2);
\end{math}
therefore, $\zeta$ is convex when $x \geq \exp\{ \vartheta + 1 -\tau^2 \}$ and concave otherwise.
\end{remark}

\citet{You-2021} considers an IMH step \citep[Section~7.4]{Robert-Casella-2004} with proposal $\sigma_i^{2*} \sim \text{IG}(\kappa_i, \lambda_i)$ to draw from \eqref{eqn:target}. Given a chain $\sigma_i^{2(1)} \ldots, \sigma_i^{2(r)}$ of $r$ draws from initial value $\sigma_i^{2(0)}$, the probability of accepting the proposed $\sigma_i^{2*}$ is
\begin{align}
\varpi_i^{(r)}
&= \min\left\{1, \frac{
f_\text{IG}(\sigma_i^{2*} \mid \kappa_i, \lambda_i)
f_\text{LN}(\sigma_i^{2*} \mid \mu_i, \tau^2)
f_\text{IG}(\sigma_i^{2(r)} \mid \kappa_i, \lambda_i)
}{
f_\text{IG}(\sigma_i^{2(r)} \mid \kappa_i, \lambda_i)
f_\text{LN}(\sigma_i^{2(r)} \mid \mu_i, \tau^2)
f_\text{IG}(\sigma_i^{2*} \mid \kappa_i, \lambda_i)
} \right\} \nonumber \\
&= \min\left\{ 1,
\exp\left\{
-\frac{1}{2 \tau^2} (\log \sigma_i^{2*} - \mu_i)^2
+\frac{1}{2 \tau^2} (\log \sigma_i^{2(r)} - \mu_i)^2
-\log \sigma_i^{2*} + \log \sigma_i^{2(r)}
\right\}
\right\};
\label{eqn:metro-ratio}
\end{align}
therefore, element $\sigma_i^{2(r+1)}$ of the chain is taken to be $\sigma_i^{2*}$ with probability $\varpi_i^{(r)}$ and $\sigma_i^{2(r)}$ with probability $1 - \varpi_i^{(r)}$. This is an independent Metropolis-Hastings step because the proposed $\sigma_i^{2*}$ is drawn independently of the previous iterate $\sigma_i^{2(r)}$. Note that use of an inverse gamma proposal yields a simplified ratio in \eqref{eqn:metro-ratio}, utilizing the form of \eqref{eqn:target} as a weighted inverse gamma density.

It is seen in Section~\ref{sec:saipe-analysis} that mixing from this sampler is not completely adequate with the SAIPE data from Section~\ref{sec:saipe-data}. In particular, some elements of $\vec{\sigma}^2$ are seen to mix very poorly. To address this, we next consider a self-tuned variant of VWS to draw exactly from conditional \eqref{eqn:target} which will then be used within the Gibbs sampler of Algorithm~\ref{alg:you}.

\begin{remark}
\label{remark:imh-mix}
Let us briefly illustrate one case of poor mixing with IMH. Kenedy County in Texas---corresponding to index $i = \text{2,655}$ in Figure~\ref{fig:saipe-trace}---has the second smallest population (343) according to the PEP data, with $y_i = 1.10$, $s_i^2 = 1.03$, $d_i = 2.74$, $\vec{z}_i = (1, 4.06)^\top$ using the formulation in Section~\ref{sec:saipe-analysis}. Suppose $\vartheta_i = y_i$ and the remaining parameters are fixed to the posterior means given in Table~\ref{tab:saipe-summary}: $\vec{\gamma} = (1.8439, -0.9024)^\top$ and $\tau^2 = 0.0953$. Then $\mu_i = \vec{z}_i^\top \vec{\gamma} = -1.8267$, $\kappa_i = 0.8708$, and $\lambda_i = 1.4072$. Figure~\ref{fig:imh} displays the resulting $\text{LN}(\mu_i, \tau^2)$ distribution which appears in the acceptance ratio and $\text{IG}(\kappa_i, \lambda_i)$ which becomes the proposal. The 99.9\% quantile of $\text{LN}(\mu_i, \tau^2)$ is $0.418$, but the probability of proposing a $\sigma^{2*}$ greater than $0.418$ is about 98.6\%. Here, IMH is very unlikely to propose draws which will be accepted when $\sigma^{2(r)}$ corresponds to a higher density region of $\text{LN}(\mu_i, \tau^2)$.
\end{remark}

\section{Self-Tuned VWS}
\label{sec:self-tune-vws}

To formulate a self-tuned variant of VWS, let us briefly recall a construction described by \citet*{vws-2026}. Suppose our objective is to draw variates from a weighted target distribution $f(x) = w(x) g(x) / \psi$ whose support is $\Omega \subseteq \mathbb{R}$. We partition $\Omega$ into regions $\mathscr{D}_j = (\alpha_{j-1}, \alpha_j]$, $j = 1, \ldots, N$, using knots $\alpha_0 < \cdots < \alpha_N$, with $\alpha_0 \equiv 0$ and $\alpha_N \equiv \infty$. Let
\begin{displaymath}
\overline{w}(x) = \sum_{j=1}^N \overline{w}_j \ind\{x \in (\alpha_{j-1}, \alpha_j]\}
\quad \text{and} \quad
\underline{w}(x) = \sum_{j=1}^N \underline{w}_j \ind\{x \in (\alpha_{j-1}, \alpha_j]\}
\end{displaymath}
be a majorizer and minorizer of $w$, respectively, so that $w(x) \leq \overline{w}_j$ and $\underline{w}_j \leq w(x)$ for $x \in (\alpha_{j-1}, \alpha_j]$. Because the majorizer and minorizer of $w$ are piecewise constants in this formulation, \citet*{vws-2026} refer to this proposal construction as ``constant VWS''.

Define
\begin{align*}
&\overline{\xi}_j = 
\int_{\mathscr{D}_j} \overline{w}_j(x) g(x) dx =
\overline{w}_j \Prob(\alpha_{j-1} < T \leq \alpha_j),
\quad \text{and} \quad \\
&\underline{\xi}_j = 
\int_{\mathscr{D}_j} \underline{w}_j(x) g(x) dx
= \underline{w}_j \Prob(\alpha_{j-1} < T \leq \alpha_j),
\end{align*}
where $T \sim g$. The density $h(x) = \sum_{j=1}^N \pi_j g_j(x)$ is a finite mixture with mixing weights $\pi_j = \overline{\xi}_j / \sum_{\ell=1}^N \overline{\xi}_\ell$ and component densities $g_j(x) \propto g(x) \ind(\alpha_{j-1} < x \leq \alpha_j)$. The function $h_0(x) = \sum_{j=1}^N \overline{w}_j g(x)$ is an unnormalized form of the density so that $h(x) \equiv h_0(x) / \psi_N$ with $\psi_N = \sum_{j=1}^N \overline{\xi}_j$. A draw from finite mixture $h$ may be accomplished in two steps: generate a label $j$ from the discrete distribution with values $1, \ldots, N$ and probabilities $\pi_1, \ldots, \pi_N$, then draw from density $g_j$. Rejection sampling is achieved by drawing $x$ from $h$ and $u$ from $\text{Uniform}(0,1)$, and accepting $x$ if $u < w(x) g(x) / h_0(x)$; otherwise, $x$ and $u$ are rejected and the process may be repeated by taking another draw. The probability of rejecting a proposed draw is $1 - \psi / \psi_N$; when $\psi$ is not readily computed, we can consider an upper bound for the probability,
\begin{equation}
\rho_{+} = \textstyle 1 - \left\{ \sum_{j=1}^N \underline{\xi}_j \right\} / \left\{ \sum_{j=1}^N \overline{\xi}_j \right\}.
\label{eqn:bound}
\end{equation}
The quantity
\begin{math}
\rho_j = \{ \overline{\xi}_j - \underline{\xi}_j \} / \sum_{\ell=1}^N \overline{\xi}_\ell
\end{math}
can be regarded as the contribution of the $j$th region to $\rho_{+}$ in the sense that $\sum_{j=1}^N \rho_j = \rho_{+}$. A $\rho_j$ which is very small relative to $\rho_+$ indicates that region $j$ contributes little to controlling the rejection rate; therefore, knots $\alpha_{j-1}$ or $\alpha_j$ may be put to better use elsewhere.

The VWS sampler just described may be applied within a Gibbs sampler. Each time the conditional is encountered, we can construct a proposal and refine it using a knot selection rule---such as Algorithm~1 in \citet*{vws-2026}---until bound $\rho_{+}$ is sufficiently reduced, then proceed with sample generation. However, this will incur a large computational overhead when used in a setting such as Algorithm~\ref{alg:you}, where it would be repeated over $N$ regions for each of $m$ subjects and $R$ iterations Gibbs sampling. Let us instead consider a self-tuned approach based on the following notion: as the overall MCMC chain evolves, the support for a particular conditional likely will not change drastically. Regions which have been useful in past iterations are likely to be useful in current iterations.

Let $h$ be a VWS proposal targeting one conditional of the Gibbs sampler; this proposal will persist for the duration of Gibbs sampling. Suppose $h$ is initially constructed with a set of internal knots $\{ \alpha_1, \ldots, \alpha_{N-1} \}$ yielding a partition of $\Omega$ with $N$ regions; for the remainder of this work, we will assume that the set of knots is empty initially. The partition will be refined and/or coarsened to achieve an adequately low rejection rate---with $\rho_+$ sufficiently below 1---so that all regions in the partition contribute substantially to $\rho_+$. Retaining regions in $h$ with low contributions is wasteful when their removal does not substantially increase the overall rejection rate. Such regions will appear in computations to draw from $h$ in the manner we have mentioned. Computational effort must also be kept under control during the tuning process; e.g., optimization routines are avoided. 

Rejection of a proposed draw during sampling will trigger consideration of a change to the partition: the rejected draw may be added as a new knot; furthermore, existing regions are reviewed to ensure that they contribute substantially. More specifically, define tolerances $\epsilon_1$ and $\epsilon_2$ which take values in $[0,1]$. A rejected draw $x$ is added as a knot if $\rho_{+}$ is not smaller than tolerance $\epsilon_1$, resulting in a refined partition. If $\rho_{+} < \epsilon_1$, regions with $\rho_j < \epsilon_2$ are potentially merged with an adjacent region by removing a knot; this is done only if it would not increase $\rho_{+}$ beyond $\epsilon_1$. Therefore, $\epsilon_1$ and $\epsilon_2$ represent thresholds for a sufficiently small rejection probability and a low contribution, respectively. In our implementation, a knot cannot be removed immediately after being added. We anticipate that $\epsilon_1$ need not be very small in a Gibbs sampler, where proposal maintenance may be more time consuming than allowing a moderate number of rejections. Furthermore, aggressive merging with a larger $\epsilon_2$ may be counterproductive if it coarsens regions which are useful as the conditional distribution evolves in the Gibbs sampler. Algorithm~\ref{alg:self-tuned-vws} details the approach just described to generate a single draw from $f$.

The refine and merge steps proposed in Algorithm~\ref{alg:self-tuned-vws} are related to established methods. The ARS and ARMS algorithms make use of rejected draws to refine the proposal. CARS avoids an excessive number of support points by keeping it fixed to a given value $N$. A rejected draw replaces the closest support point if the swap improves the rejection rate. We find it more intuitive to specify probabilities $\epsilon_1$ and $\epsilon_2$ than an $N$, especially because there are $m$ proposals where some may require very few support points while others may benefit from a larger number. Furthermore, it can be beneficial to use many support points in small regions where the density is varying rapidly. We have not included CARS in our computational studies for these reasons. We have also not included FUSS, which is presented as a framework rather than a specific algorithm. VWS may be considered among the methods of proposal construction in FUSS, which otherwise are not assumed to be proper envelopes for rejection sampling. In FUSS, the strategy is to begin with a large number of support points and prune the excessive ones; however, only brief consideration by \citet{Martino-EtAl-2015} is given to curating this set between Gibbs iterations. Several possible pruning approaches are described therein; with our merge step closely following their Algorithm~P4. \citet{Martino-EtAl-2015} sequentially remove support points to minimize the increase in L1 distance between the proposal and target, until all low-contribution points have been pruned. The use of L1 distance and the conceptualization of low contribution regions are closely related to the use of $\rho_+$ and $\rho_j$ in the present work. In contrast, we avoid the computational burden of searching for the minimal increase at each step and simply merge low-contribution regions in the order we encounter them.

\begin{algorithm}
\caption{Generate a single draw from $f$ using self-tuned VWS with tolerances $\epsilon_1$ and $\epsilon_2$.}
\label{alg:self-tuned-vws}
\begin{algorithmic}[1]
\Do
\State Draw $u$ from $\text{Uniform}(0,1)$ and $x$ from $h$. \Comment{Rejection sampling}
\State $\text{Accept} \gets \ind\{ u \leq f_0(x) / h_0(x) \}$ where $f_0(x) = w(x) g(x)$ and $h_0(x) = \overline{w}(x) g(x)$.
\State Let $N$ be the current number of knots in $h$.
\State Let $\alpha_0, \ldots, \alpha_N$ be the current knots of $h$.
\State Let $\rho_{+}$ be current value of bound \eqref{eqn:bound}.
\State Let $\rho_1, \ldots, \rho_N$ be the current contribution of each region.
\If{$\text{Accept} = 0$ and $\rho_{+} < \epsilon_1$} \Comment{Merge regions}
\For{$j = 1, \ldots, N-1$}
\If{$\rho_j < \epsilon_2$}
\State Let $\tilde{h}$ be equal to $h$ with knot $\alpha_j$ excluded.
\State Let $\tilde{\rho}_+$ be the bound \eqref{eqn:bound} for $\tilde{h}$.
\State Assign $h$ to $\tilde{h}$ if $\tilde{\rho}_+ < \epsilon_1$.
\EndIf
\EndFor
\ElsIf{$\text{Accept} = 0$ and $\rho_{+} \geq \epsilon_1$} \Comment{Refine regions}
\State Add $x$ as a knot in $h$.
\EndIf
\DoWhile{$\text{Accept} = 0$}
\State \Return $x$ and $h$.
\end{algorithmic}
\end{algorithm}

\section{VWS within Gibbs for Joint SAE Model}
\label{sec:vws-within-gibbs}

To address the mixing issues discussed in Section~\ref{sec:you}, we consider exact draws from conditional \eqref{eqn:target} with self-tuned VWS. It can be used to effectively construct a proposal for the conditionals \eqref{eqn:target}, which are not log-concave. The criteria $\epsilon_1$ and $\epsilon_2$ may be selected to balance effort to reduce the bound $\rho_+$ with time spent refining the proposal so that steady progress can be made in Gibbs sampling.

We first obtain terms in the finite mixture that will become the proposal distribution. Conditional~\eqref{eqn:target} may be viewed as a weighted density with
\begin{align}
f(\sigma_i^2) \propto
\underbrace{f_\text{IG}(\sigma_i^2 \mid \kappa_i, \lambda_i)}_{w(\sigma_i^2)}
\underbrace{f_\text{LN}(\sigma_i^2 \mid \mu_i, \tau^2)}_{g(\sigma_i^2)}.
\label{eqn:target-tx}
\end{align}
The distribution $\text{IG}(\sigma_i^2 \mid \kappa_i, \lambda_i)$ is unimodal with mode $\eta_i = \lambda_i / (\kappa_i + 1)$. Therefore, the respective majorizer and minorizer of $w$ are
\begin{displaymath}
\overline{w}(x) = \sum_{j=1}^N \overline{w}_j \ind(\alpha_{j-1} < x \leq \alpha_j),
\quad
\underline{w}(x) = \sum_{j=1}^N \underline{w}_j \ind(\alpha_{j-1} < x \leq \alpha_j),
\end{displaymath}
with
\begin{align*}
\overline{w}_j =
\begin{cases}
w(\alpha_{j-1}), & \text{if $\eta_i \leq \alpha_{j-1}$}, \\
w(\alpha_j), & \text{if $\eta_i > \alpha_j$}, \\
w(\eta_i), & \text{otherwise},
\end{cases}
%
\quad \text{and} \quad
%
\underline{w}_j =
\begin{cases}
w(\alpha_j), & \text{if $\eta_i \leq \alpha_{j-1}$}, \\
w(\alpha_{j-1}), & \text{if $\eta_i > \alpha_j$}, \\
\min\{ w(\alpha_j), w(\alpha_{j-1}) \}, & \text{otherwise}.
\end{cases}
\end{align*}
The resulting VWS proposal is the finite mixture $h(x) = \sum_{j=1}^N \pi_j g_j(x)$ whose mixing weights $\pi_j$ are proportional to
\begin{math}
\overline{\xi}_j = \alpha_{j-1} \Prob(\alpha_{j-1} < T_i \leq \alpha_j)
\end{math}
for $T_i \sim \text{LN}(\mu_i, \tau^2)$ and components $g_j$ are densities of $\text{LN}(\mu_i, \tau^2)$ truncated to $(\alpha_{j-1}, \alpha_j]$. The bound $\rho_{+}$ is computed from $\overline{\xi}_j$ and
\begin{math}
\underline{\xi}_j = \alpha_j \Prob(\alpha_{j-1} < T_i \leq \alpha_j).
\end{math}

Algorithm~\ref{alg:self-tuned-vws} can be used with the expressions in this section to implement self-tuned VWS for a single conditional of the Gibbs sampler in Section~\ref{sec:you}. This can be repeated for each conditional by maintaining $m$ proposals $h_1, \ldots, h_m$. A complete self-tuned VWS-within-Gibbs sampler is presented in Algorithm~\ref{alg:stvws-within-gibbs}. As in Section~\ref{sec:self-tune-vws}, we assume the proposals are each initialized from a single region $\mathscr{D}_1 = (0, \infty]$.

\begin{remark}
\label{remark:mem}
If the $i$th proposal contains $N_i^{(r)}$ regions after the $r$th iteration of the Gibbs sampler, memory usage will be of the order
\begin{math}
\max\limits_{r = 1 \ldots, R} \sum_{i=1}^m \sum_{j=1}^{N_i^{(r)}} B,
\end{math}
where $B$ bytes are required to store a region's contents. In our implementation, such contents include variables such as the endpoints $(\alpha_{j-1}, \alpha_j)$, the optimal values $(\overline{w}_j, \underline{w}_j)$. and several functions including $w$ as well as the density, CDF, and quantile function for $g$.
\end{remark}

\begin{algorithm}
\caption{Self-tuned VWS for joint SAE model with tolerances $\epsilon_1$ and $\epsilon_2$. }
\label{alg:stvws-within-gibbs}
\begin{algorithmic}[1]
\For{$r = 1, \ldots, R$}
\State Draw $\vec{\vartheta}$ and $\vec{\theta} = (\vec{\beta}, \vec{\gamma}, \phi^2, \tau^2)$ using steps 1--5 in Algorithm~\ref{alg:you}.
\For{$i = 1, \ldots, m$}
\State Let $\mu_i = \vec{z}_i^\top \vec{\gamma}$, $\kappa_i = (d_i-1)/2$ and $\lambda_i = (y_i - \vartheta_i)^2 / 2 + d_i s_i^2 / 2$ for target \eqref{eqn:target-tx}.
\State Draw $\sigma_i^2$ using Algorithm~\ref{alg:self-tuned-vws} with proposal $h_i$ and expressions in Section~\ref{sec:vws-within-gibbs}.
\State Update $h_i$ with modifications from tuning.
\EndFor
\State Let $\vec{\vartheta}^{(r)}$, $\vec{\sigma}^{2(r)}$, and $\vec{\theta}^{(r)}$ be respective draws from $r$th iteration.
\EndFor
\end{algorithmic}
\end{algorithm}

\section{Simulations}
\label{sec:sim}

Several simulation studies are now presented to compare the performance of the IMH and VWS samplers in the context of the model in Section~\ref{sec:you}. Section~\ref{sec:sim-cond} considers drawing from several specific conditional distributions of the form \eqref{eqn:target-tx} while Section~\ref{sec:sim-post} considers the complete Gibbs sampler for the joint SAE model. Mixing of chains is assessed primarily by effective sample size (ESS) via the \texttt{mcmcse} R package \citep*{mcmcse}: the functions \texttt{ess} and \texttt{multiESS} are used to compute univariate and multivariate ESS, respectively. ESS values are rounded to integers, with larger values indicating better mixing. ESS per second is also reported in some results to serve as a measure of throughput; it is obtained by dividing ESS by the elapsed time.

\subsection{Conditional Simulation}
\label{sec:sim-cond}

This section illustrates the IMH and self-tuned VWS samplers from Sections~\ref{sec:you} and \ref{sec:vws-within-gibbs} to generate draws from a single conditional of the form \eqref{eqn:target-tx}. The parameters $\mu = 0$ and $\lambda = 1$ are fixed in this study. We vary $\kappa \in \{ 10, 50 \}$ which corresponds to a smaller and larger degrees of freedom. We also vary $\tau \in \{ 0.5, 1.0 \}$ which corresponds to higher and lower precision of the regression model for the sampling variance.

\begin{figure}
\centering
\caption{Trace plots of draws 199,001 to 200,000 from \eqref{eqn:target-tx} for $\sigma^2$ using IMH under four parameter settings. (IMH: independent Metropolis-Hastings.)}
\label{fig:cond-metro}
\begin{subfigure}[t]{0.35\textwidth}
\centering
\includegraphics[width=1.0\textwidth]{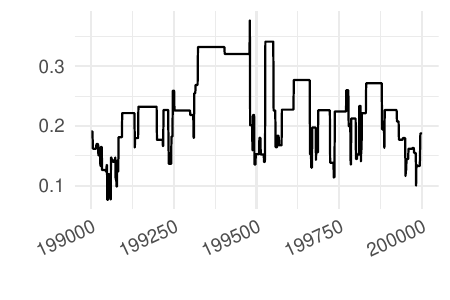}
\caption{$\kappa = 10, \tau = 0.5.$}
\label{fig:cond-metro-1-1}
\end{subfigure}
\begin{subfigure}[t]{0.35\textwidth}
\centering
\includegraphics[width=1.0\textwidth]{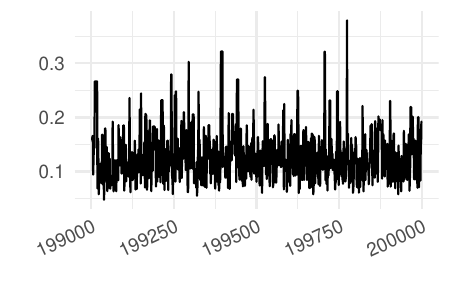}
\caption{$\kappa = 10, \tau = 1.0.$}
\label{fig:cond-metro-1-2}
\end{subfigure}
\begin{subfigure}[t]{0.35\textwidth}
\centering
\includegraphics[width=1.0\textwidth]{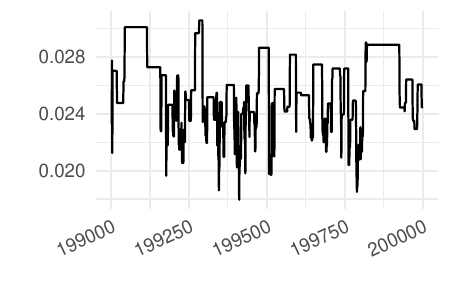}
\caption{$\kappa = 50, \tau = 0.5.$}
\label{fig:cond-metro-2-1}
\end{subfigure}
\begin{subfigure}[t]{0.35\textwidth}
\centering
\includegraphics[width=1.0\textwidth]{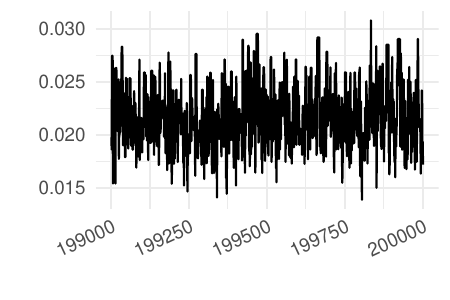}
\caption{$\kappa = 50, \tau = 1.0.$}
\label{fig:cond-metro-2-2}
\end{subfigure}
\end{figure}

First consider chains from the IMH algorithm using \eqref{eqn:metro-ratio} which have been initialized with the maximizer of \eqref{eqn:target-tx} and run for 200,000 steps. Figure~\ref{fig:cond-metro} displays four such chains corresponding to the chosen levels of the parameters. Mixing is seen to be notably slower under $\tau = 0.5$. Table~\ref{tab:cond-metro} provides summary statistics to accompany Figure~\ref{fig:cond-metro}. Here, ESS is computed on the entire chain of length $\text{200,000}$. The number of rejections and sample autocorrelation at lag 1 are also displayed. These metrics confirm that mixing is adequate in the cases that $\tau = 1.0$ but questionable with $\tau = 0.5$. In the latter cases, we notice a high autocorrelation and a larger proportion of the steps being rejections where the chain does not move.

Self-tuned VWS is also considered for the same levels of $(\kappa, \tau)$, additionally varying $\epsilon_1 \in \{ 0.50, 0.75 \}$ and $\epsilon_2 \in \{ 0.001, 0.01 \}$. Here we generate $n = 20$ independent observations from each proposal starting with no internal knots. The sequence of $n$ draws is repeated for $R = \text{10,000}$ repetitions to express variability in tuning which depends on the sequence of proposed draws. Figure~\ref{fig:cond-vws} shows that tolerance $\epsilon_1 = 0.75$ tends to be satisfied within 20 iterations for all four parameter settings; however, several additional iterations are needed to achieve $\epsilon_1 = 0.5$. The number of internal knots settles to under 25 except in the cases where $\kappa = 50$ and $\tau = 0.5$; here, the knot count approaches 45.

Table~\ref{tab:cond-vws} summarizes elapsed times and rejection counts; both quantities represent sums over the $R$ repetitions. Recall that we produce a large number ($R = \text{10,000}$) of short chains ($n = 20$) where the underlying target distribution is not changing. Here it is useful to utilize the smaller rejection tolerance of $\epsilon_1 = 0.50$. The setting of $\epsilon_2$ is most consequential when $\kappa = 50$, $\tau = 0.5$, and $\epsilon_1 = 0.75$. Here the proposal appears to accumulate a number of low contribution rejections so that it is beneficial to merge them more proactively. Otherwise, the selection of $\epsilon_2$ may be more impactful when longer chains are under consideration.

\begin{table}
\centering
\caption{Summary of chains shown in Figure~\ref{fig:cond-metro} where $\rho(1)$ is sample autocorrelation at lag 1. (ESS: effective sample size.)}
\label{tab:cond-metro}
\begin{tabular}{rrrrr}
\toprule
\multicolumn{1}{c}{$\kappa$} &
\multicolumn{1}{c}{$\tau$} &
\multicolumn{1}{c}{ESS} &
\multicolumn{1}{c}{Rejections} &
\multicolumn{1}{c}{$\rho(1)$} \\
\midrule
10 & 0.5 &  1,165 & 175,247 & 0.974 \\
10 & 1.0 & 78,798 &  42,387 & 0.409 \\
50 & 0.5 &    337 & 174,174 & 0.974 \\
50 & 1.0 & 63,774 &  45,895 & 0.439 \\
\bottomrule
\end{tabular}
\end{table}

\begin{table}
\centering
\caption{Summary of results for self-tuned VWS over $R = \text{10,000}$ repetitions of $n = 20$ draws. (VWS: vertical weighted strips.)}
\label{tab:cond-vws}
\begin{subtable}[t]{\textwidth}
\centering
\caption{Sum of elapsed times in seconds.}
\label{tab:cond-vws-elapsed}
\begin{tabular}{rrrrrr}
\toprule
\multicolumn{2}{c}{} &
\multicolumn{2}{c}{$\epsilon_1 = 0.50$} &
\multicolumn{2}{c}{$\epsilon_1 = 0.75$} \\
\cmidrule(lr){3-4}
\cmidrule(lr){5-6}
\multicolumn{1}{c}{$\kappa$} &
\multicolumn{1}{c}{$\tau$} &
\multicolumn{1}{c}{$\epsilon_2 = 0.001$} &
\multicolumn{1}{c}{$\epsilon_2 = 0.01$} &
\multicolumn{1}{c}{$\epsilon_2 = 0.001$} &
\multicolumn{1}{c}{$\epsilon_2 = 0.01$} \\
\cmidrule(lr){1-1}
\cmidrule(lr){2-2}
\cmidrule(lr){3-3}
\cmidrule(lr){4-4}
\cmidrule(lr){5-5}
\cmidrule(lr){6-6}
10 & 0.5 & 10.78 & 12.12 & 10.51 & 12.71 \\
10 & 1.0 &  6.10 &  6.53 &  6.41 &  7.25 \\
50 & 0.5 & 27.00 & 28.81 & 76.07 & 66.99 \\
50 & 1.0 &  9.93 & 10.10 & 15.16 & 15.96 \\
\bottomrule
\end{tabular}
\end{subtable}

\begin{subtable}[t]{\textwidth}
\centering
\caption{Sum of rejection counts.}
\label{tab:cond-vws-rejections}
\begin{tabular}{rrrrrr}
\toprule
\multicolumn{2}{c}{} &
\multicolumn{2}{c}{$\epsilon_1 = 0.50$} &
\multicolumn{2}{c}{$\epsilon_1 = 0.75$} \\
\cmidrule(lr){3-4}
\cmidrule(lr){5-6}
\multicolumn{1}{c}{$\kappa$} &
\multicolumn{1}{c}{$\tau$} &
\multicolumn{1}{c}{$\epsilon_2 = 0.001$} &
\multicolumn{1}{c}{$\epsilon_2 = 0.01$} &
\multicolumn{1}{c}{$\epsilon_2 = 0.001$} &
\multicolumn{1}{c}{$\epsilon_2 = 0.01$} \\
\cmidrule(lr){1-1}
\cmidrule(lr){2-2}
\cmidrule(lr){3-3}
\cmidrule(lr){4-4}
\cmidrule(lr){5-5}
\cmidrule(lr){6-6}
10 & 0.5 & 247,565 & 249,866 & 302,007 & 311,712 \\
10 & 1.0 & 155,091 & 156,390 & 170,169 & 174,325 \\
50 & 0.5 & 461,697 & 463,834 & 505,715 & 527,268 \\
50 & 1.0 & 218,978 & 220,070 & 234,855 & 238,899 \\
\bottomrule
\end{tabular}
\end{subtable}
\end{table}

\subsection{Posterior Simulation}
\label{sec:sim-post}

This section compares IMH from Section~\ref{sec:you} with VWS from Section~\ref{sec:vws-within-gibbs} within Algorithm~\ref{alg:you}. We consider the mixing and computational performance of the overall MCMC in a simulation setting. The simulation proceeds by generating a dataset from model \eqref{eqn:model} under one of several settings. Two levels for the number of small areas $m \in \{ 500, \text{2,000} \}$ are considered. Sample sizes $n_i$ for the areas are generated from a $\chi_{16}^2$ distribution. The design matrix $\vec{Z}$ consists of two columns: an intercept and the vector with elements $\log n_1, \ldots, \log n_m$. The associated coefficient is taken to be $\vec{\gamma} = (2.6, -1)^\top$. Additionally, three levels of each tolerance $\epsilon_1 \in \{ 0.5, 0.75, 0.85 \}$ and $\epsilon_2 \in \{ 0.0001, 0.001, 0.01 \}$ are considered for VWS. Data-generating values of the remaining parameters are taken to be $\tau^2 = 0.25$, $\phi^2 = 0.2$, and $\vec{\beta} = (1.5, 0.85)^\top$. The matrix $\vec{X}$ is assumed to have two columns: an intercept and a vector generated from $\text{N}(8, 2^2)$. Degrees of freedom $d_i$ are taken to be $n_i - 1$.

The simulation consists of 18 levels combining variations of $m$, $\epsilon_1$, and $\epsilon_2$; for each level, $S = 500$ datasets are generated and each of the MCMC methods are applied. For VWS, a chain of length 3,000 is produced where the first 1,000 are discarded as burn-in and the remaining $R = \text{2,000}$ are saved to summarize the posterior. To accommodate slower mixing suspected under IMH, its chains are run to length 30,000; the first 28,000 of which are discarded as burn-in and the remaining $R = \text{2,000}$ are saved. Initial values for $\vec{\beta}$ and $\phi^2$ are obtained by fitting an ordinary least squares regression of $\vec{y}$ on $\vec{X}$; similarly, initial $\vec{\gamma}$ and $\tau^2$ values are obtained from a regression of $\log \vec{s}^2 = (\log s_1^2, \ldots, \log s_m^2)^\top$ on $\vec{Z}$. Initial value $\vec{\sigma}^2 = (1, \ldots, 1)^\top$ is taken for the latent variances. No initial value is required for the vector of latent means $\vec{\vartheta}$ as it is drawn as the first step in the Gibbs sampler.

Results from the simulations are summarized in Table~\ref{tab:sim-post}. For both MCMC samplers, the elapsed time in seconds is measured for each of the $S$ simulated datasets and averaged. To capture mixing of the chains for $\sigma_1^2, \ldots, \sigma_m^2$, ESS is computed for each and summarized using the 0\% (minimum), 1\%, and 2.5\% quantiles. This is repeated for each of the $S$ datasets and the quantiles are averaged over the $S$ datasets. Rejection counts are summarized for both IMH and VWS; the number of rejections is totaled for each of the $S$ datasets and the average is recorded. The number of tuning steps from VWS is similarly recorded: the overall number tuning steps (refinements and merges) is totaled for both the burn-in period and afterward; each count is averaged over the $S$ datasets. Because $\epsilon_1$ and $\epsilon_2$ are not relevant to IMH, results from IMH are also averaged over the nine levels of these tolerances. Table~\ref{tab:sim-post-ext} presents several additional results: ESS per second for $\vec{\sigma}^2$ and ESS of $\vec{\vartheta}$ are summarized by the 0\%, 1\%, and 2.5\% quantiles in the same way as ESS for $\vec{\sigma}^2$, and multivariate ESS is displayed for parameter $\vec{\theta}$.

As anticipated, IMH is substantially faster than VWS per iteration because of its computational simplicity. The rejection rate for IMH over all proposed draws is
\begin{math}
\red{
\text{5,501,678} / (\text{30,000} m) \approx 36.7\%
}
\end{math}
for $m = 500$ and
\begin{math}
\red{
\text{22,059,787} / (\text{30,000} m) \approx 36.77\%
}
\end{math}
for $m = \text{2,000}$. However, ESS indicates poor mixing for some elements in $\vec{\sigma}^2$. The worst mixing chains are seen to average about \red{36} and \red{20} ESS, respectively, for $m \in \{ 500, \text{2,000} \}$. ESS quantiles for $\vec{\vartheta}$ and multivariate ESS for $\vec{\theta}$ indicate that those elements are mixing much more adequately.

Mixing of $\vec{\sigma}^2$ elements is seen to be significantly improved under VWS. Here the worst mixing chains average an ESS of \red{1,501} or larger under $m = 500$ and \red{1,448} or larger under $m = \text{2,000}$. Therefore, drawing from the exact conditional with rejection sampling appears to yield adequate mixing in all chains. The fastest times are seen when taking $\epsilon_1$ larger to permit more rejections. The three values of $\epsilon_2$ considered in this setting perform similarly. ESS for $\vec{\vartheta}$ and $\vec{\theta}$ indicate mixing under VWS that is at least on par with IMH or slightly better, as anticipated.

\begin{table}
\centering
\caption{Summary of results from posterior simulation. Metrics are averaged over $S = 500$ simulation repetitions. Elapsed times are displayed in seconds. (ESS: effective sample size; VWS: vertical weighted strips; IMH: independent Metropolis-Hastings.)}
\label{tab:sim-post}
\begin{tabular}{rrlrrrrrrr}
\toprule
\multicolumn{3}{l}{(a) IMH} &
\multicolumn{3}{c}{ESS $\vec{\sigma}^2$} &
\multicolumn{2}{c}{} &
\multicolumn{2}{c}{} \\
\cmidrule(lr){4-6}
\multicolumn{1}{c}{$m$} &
\multicolumn{1}{c}{} &
\multicolumn{1}{c}{} &
\multicolumn{1}{c}{Min} &
\multicolumn{1}{c}{1\%} &
\multicolumn{1}{c}{2.5\%} &
\multicolumn{1}{c}{Elapsed} &
\multicolumn{1}{c}{Rejections} &
\multicolumn{1}{c}{} &
\multicolumn{1}{c}{} \\
\cmidrule(lr){1-3}
\cmidrule(lr){4-6}
\cmidrule(lr){7-7}
\cmidrule(lr){8-8}
  500 &  &  & 36 & 83 & 122 &  7.87 &  5,501,678 &  & \\
2,000 &  &  & 20 & 78 & 118 & 28.80 & 22,059,787 &  & \\
\midrule
\multicolumn{3}{l}{(b) VWS} &
\multicolumn{3}{c}{ESS $\vec{\sigma}^2$} &
\multicolumn{2}{c}{} &
\multicolumn{2}{c}{Tuning Steps} \\
\cmidrule(lr){4-6}
\cmidrule(lr){9-10}
\multicolumn{1}{c}{$m$} &
\multicolumn{1}{c}{$\epsilon_1$} &
\multicolumn{1}{c}{$\epsilon_2$} &
\multicolumn{1}{c}{Min} &
\multicolumn{1}{c}{1\%} &
\multicolumn{1}{c}{2.5\%} &
\multicolumn{1}{c}{Elapsed} &
\multicolumn{1}{c}{Rejections} &
\multicolumn{1}{c}{Burn} &
\multicolumn{1}{c}{Keep} \\
\cmidrule(lr){1-3}
\cmidrule(lr){4-6}
\cmidrule(lr){7-7}
\cmidrule(lr){8-8}
\cmidrule(lr){9-10}
  500 & 0.50 & 0.0001 & 1,505 & 1,722 & 1,783 &  32.11 &   310,772 &  6,512 &    270 \\
  500 & 0.50 & 0.001  & 1,529 & 1,722 & 1,784 &  30.27 &   316,404 &  7,789 &    760 \\
  500 & 0.50 & 0.01   & 1,501 & 1,720 & 1,782 &  31.98 &   355,669 & 23,397 & 18,539 \\
  500 & 0.75 & 0.0001 & 1,512 & 1,720 & 1,782 &  23.60 &   491,600 &  3,756 &    120 \\
  500 & 0.75 & 0.001  & 1,538 & 1,722 & 1,783 &  23.56 &   490,909 &  4,199 &    230 \\
  500 & 0.75 & 0.01   & 1,516 & 1,718 & 1,780 &  23.98 &   510,147 &  7,563 &  2,212 \\
  500 & 0.85 & 0.0001 & 1,520 & 1,724 & 1,786 &  21.81 &   596,636 &  2,970 &     86 \\
  500 & 0.85 & 0.001  & 1,511 & 1,722 & 1,784 &  21.90 &   591,884 &  3,287 &    158 \\
  500 & 0.85 & 0.01   & 1,522 & 1,724 & 1,787 &  21.82 &   600,818 &  5,314 &    996 \\
2,000 & 0.50 & 0.0001 & 1,469 & 1,745 & 1,804 & 128.45 & 1,253,760 & 25,467 &    896 \\
2,000 & 0.50 & 0.001  & 1,448 & 1,744 & 1,804 & 124.70 & 1,272,453 & 30,064 &  2,323 \\
2,000 & 0.50 & 0.01   & 1,462 & 1,745 & 1,804 & 127.66 & 1,424,733 & 83,260 & 57,214 \\
2,000 & 0.75 & 0.0001 & 1,462 & 1,744 & 1,804 &  97.58 & 1,985,627 & 14,692 &    395 \\
2,000 & 0.75 & 0.001  & 1,478 & 1,745 & 1,806 &  98.19 & 1,990,311 & 16,266 &    701 \\
2,000 & 0.75 & 0.01   & 1,467 & 1,746 & 1,806 &  97.71 & 2,052,085 & 27,355 &  6,210 \\
2,000 & 0.85 & 0.0001 & 1,463 & 1,744 & 1,806 &  89.82 & 2,420,090 & 11,644 &    281 \\
2,000 & 0.85 & 0.001  & 1,467 & 1,745 & 1,806 &  91.32 & 2,407,449 & 12,727 &    459 \\
2,000 & 0.85 & 0.01   & 1,475 & 1,746 & 1,806 &  90.95 & 2,428,163 & 19,517 &  2,800 \\
\bottomrule
\end{tabular}
\end{table}

\section{Application to SAIPE Dataset}
\label{sec:saipe-analysis}

We now consider several variations of the Gibbs sampler in Algorithm~\ref{alg:you} to investigate the county-level SAIPE dataset introduced in Section~\ref{sec:saipe-data}. The objectives are to determine whether mixing is substantially improved when VWS replaces the IMH step in practice and ensure that VWS does not become too computationally costly. 

For the log-count of children in poverty, we assume the regression $\vec{x}_i^\top \vec{\beta} = \beta_0 + \beta_1 \log\{1 + \text{SNAP}_i \} + \beta_2 \log\{1 + \text{PEP}_i \}$. SNAP counts are not supplied in nine counties; we assume $\text{SNAP}_i$ to be zero here (no SNAP participants) to proceed with the analysis. For the sampling variances, we assume the regression $\vec{z}_i^\top \vec{\gamma} = \gamma_0 + \gamma_1 \log n_i$ to match \citet{You-2021}. We use $d_i = 0.36 \cdot \sqrt{n_i}$ for the modeled degrees of freedom $d_i$ of $s_i^2$; this is recommended by \citet{Maples-Bell-Huang-2009} based on their study from past ACS data and an alternative variance estimate $s_i^2$ of the log-transformed point estimates that makes use of survey results internal to the U.S. Census Bureau. There are $m = \text{3,140}$ areas in the data after excluding three counties in Texas: Loving County has a point estimate of zero where the log-transform is problematic, while Crockett and King Counties have $d_i$ values less than one which yield a negative $\kappa_i = (d_i - 1) / 2$. Additional details on preparation of the analysis data are provided in the replication materials.

We consider several variations of VWS. ``VWS0'' is the standard VWS method where a new proposal is constructed when a draw of any $\sigma_i^2$ is needed. Here we utilize Algorithm~1 of \citet*{vws-2026} to refine each proposal before sampling until there are either $N = 50$ regions or $\rho_+ < \epsilon_1 \equiv 0.85$ is satisfied. ``VWS1'' is the self-tuned method presented in Algorithm~\ref{alg:self-tuned-vws}. Tuning parameters $\epsilon_1 \in \{ 0.75, 0.85 \}$ and $\epsilon_2 \in \{ 0.001, 0.01 \}$ are considered for VWS1.

A third variation ``VWS2'' is motivated by the observation that tuning adjustments in VWS1 tend to reduce dramatically and level off after an initial number of Gibbs iterations. Therefore, VWS2 is the same as VWS1 except the tuning period is limited to the first 100 Gibbs iterations. Alterations to knots are halted thereafter, but the $m$ proposals continue to be updated with current values of $\mu_i$, $\tau^2$, $\kappa_i$, and $\lambda_i$ from the MCMC chain to ensure that they remain valid envelopes for rejection sampling. The tuning period of 100 was chosen by visually inspecting the number of adjustments per iteration in VWS1 under the four combinations of $\epsilon_1$ and $\epsilon_2$. The first column of plots in Figure~\ref{fig:unmatch-tune-ext1} displays this information; not shown are initial iterations where tuning activity is higher.

We consider several additional methods for comparison. First is a component-wise adaptive Metropolis-Hastings (AMH) proposed by \citet*{Haario-Saksman-Tamminen-2005}, as presented in Section~3.1 of \citet{Roberts-Rosenthal-2009}. This variant operates on univariate components of the chain individually and is well-suited for the present application where components of $[\vec{\sigma}^2 \mid \rest]$ are independent. Here, the proposal variate is taken to be $\sigma_i^{2*} = \exp(\phi_i^*)$ with $\phi_i^* \sim \text{N}\big( \log \sigma_i^{(r)}, 2.4^2 \cdot v_i^{(r)} \big)$, where $v_i^{(r)}$ approximates the variance of $[\log \sigma_i^2 \mid \rest]$ using past draws of $\log \sigma_i^2$. We accept $\sigma_i^{2(r+1)} = \sigma_i^{2*}$ with probability
\begin{align}
\varpi_i^{(r)}
&= \min\left\{ 1, \frac{
f_\text{IG}(\exp(\phi^*) \mid \kappa_i, \lambda_i) \cdot
f_\text{LN}(\exp(\phi^*) \mid \mu_i, \tau^2) \cdot
\exp(\phi^*)
}{
f_\text{IG}(\exp(\phi^{(r)}) \mid \kappa_i, \lambda_i) \cdot
f_\text{LN}(\exp(\phi^{(r)}) \mid \mu_i, \tau^2) \cdot
\exp(\phi^{(r)})
} \right\}
\label{eqn:amh-ratio}
\end{align}
and let $\sigma_i^{2(r+1)} = \sigma_i^{2(r)}$ with probability $1 - \varpi_i^{(r)}$. The $\exp(\phi)$ term in \eqref{eqn:amh-ratio} arises from the Jacobian of the transformation from $\phi_i^*$ to $\sigma_i^{2*}$ which ensures that it is nonnegative. A second method for comparison is ARMS using the \texttt{armspp} R package \citep{armspp}. ARMS is well-known to be suitable for targets that are not log-concave, as noted in Remark~\ref{remark:concavity}, but produces an MCMC chain rather than drawing exactly from the target. Here we allow up to 100 support points to construct the proposal based on the log-density of target \eqref{eqn:target}. After drawing $\sigma_i^{2(r)}$ in the $r$th Gibbs iteration, the 5\%, 50\%, and 95\% quantiles of the proposal distribution are saved and used as initial support points in the $(r+1)$th Gibbs iteration, as suggested by \citet{Gilks-Best-Tan-1995}.

\begin{figure}
\centering
\caption{Empirical CDF of ESS for $\vec{\sigma}^2$: IMH ($\CIRCLE$), AMH ($\blacktriangle$), and VWS ($\blacksquare$). ARMS is displayed as a curve without symbols which is almost indistinguishable from VWS. (ESS: effective sample size; VWS: vertical weighted strips; IMH: independent Metropolis-Hastings.)}
\label{fig:saipe-ess}
\includegraphics[width=0.5\textwidth]{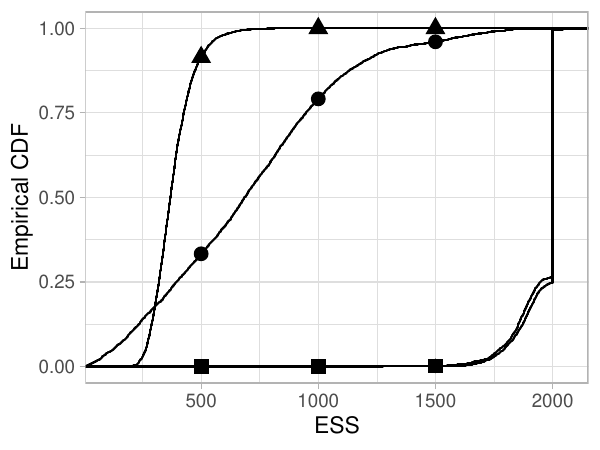}
\end{figure}

As in the simulation from Section~\ref{sec:sim}, a chain of length 30,000 is produced using IMH with the first 28,000 elements discarded as burn-in. The same settings are used with AMH. A chain of length 3,000 is produced under all other samplers with the first 1,000 elements discarded as burn-in. Therefore, 2,000 draws are saved using all samplers. Initial values are selected as in Section~\ref{sec:sim-post}. Table~\ref{tab:saipe-mixing} summarizes mixing obtained using the various samplers. Results include several quantiles of ESS and ESS per second among the chains for $\sigma_1^2, \ldots, \sigma_m^2$, elapsed times, and rejection counts. Figure~\ref{fig:saipe-ess} displays the empirical CDF of ESS for the elements of $\vec{\sigma}^2$ under each of IMH, AMH, ARMS, and VWS.

\begin{figure}
\centering
\caption{Selected saved chains for $\vec{\sigma}^2$ for VWS1 with $\epsilon_1 = 0.85$ and $\epsilon_2 = 0.01$ (light red), and IMH (dark blue) within Gibbs. The top row displays the three chains ($i = 72, 2655, 75$) with lowest ESS using IMH: Bristol Bay Borough, Alaska; Dillingham Census Area, Alaska, and Kenedy County, Texas. The bottom row shows the three chains ($i = 3139, 2387, 2739$) with lowest ESS under VWS: Teton County, Wyoming; Liberty County, Montana; and Radford City, Virginia. (ESS: effective sample size; VWS: vertical weighted strips; IMH: independent Metropolis-Hastings.)}
\label{fig:saipe-trace}
\includegraphics[width=0.32\textwidth]{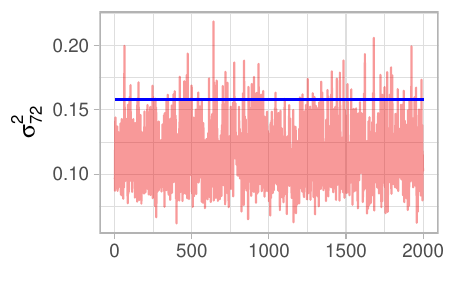}
\includegraphics[width=0.32\textwidth]{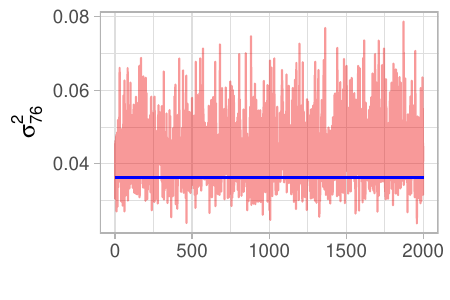}
\includegraphics[width=0.32\textwidth]{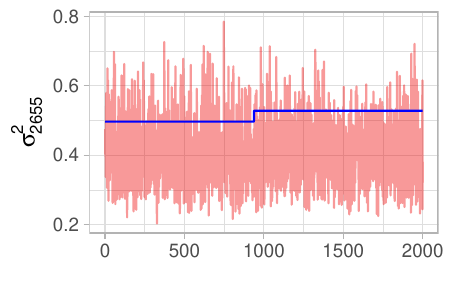}
\includegraphics[width=0.32\textwidth]{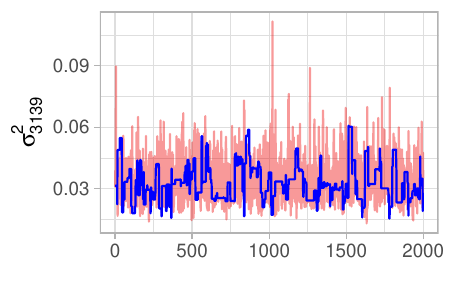}
\includegraphics[width=0.32\textwidth]{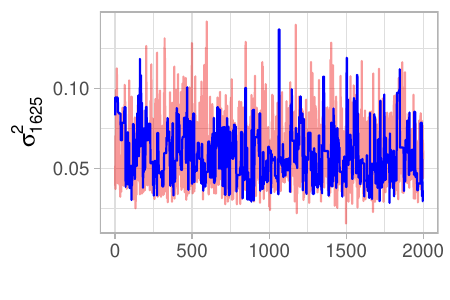}
\includegraphics[width=0.32\textwidth]{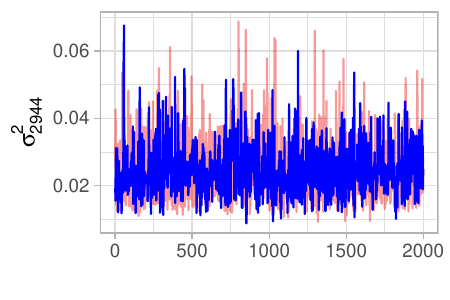}
\end{figure}

As expected, IMH iterations are faster than any VWS method but overall mixing of the $\vec{\sigma}^2$ chains is improved substantially under VWS. Figure~\ref{fig:saipe-trace} includes trace plots for the three $\vec{\sigma}^2$ elements having lowest ESS under IMH and VWS, respectively. Several chains are not able to move under IMH in its worst cases---see Remark~\ref{remark:imh-mix}---while all chains for VWS appear to be mixing well in all cases. A large variation is seen in the minimum ESS among the VWS methods; this should be due only to randomness as all VWS methods produce exact draws from distribution \eqref{eqn:target}. 

\begin{figure}
\centering
\caption{Several metrics by Gibbs iteration using self-tuned VWS with $\epsilon_1 = 0.85$, $\epsilon_2 = 0.01$. (\subref{fig:saipe-tunes-vws1-4})--(\subref{fig:saipe-rejects-vws1-4}) Number of tuning adjustments, regions, and rejections, respectively, with VWS1. (\subref{fig:saipe-rejects-vws2-4-tail}) Number of rejections under VWS2. Initial iterations are omitted from each plot to aid visualization. (VWS: vertical weighted strips.)}
\label{fig:saipe-tune}
\begin{subfigure}[t]{0.32\textwidth}
\centering
\includegraphics[width=\textwidth]{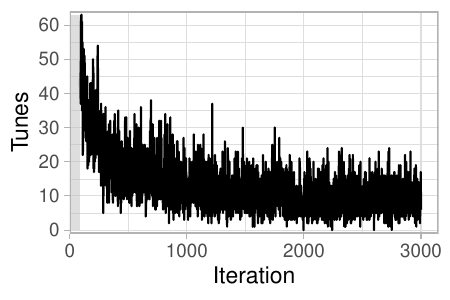}
\caption{}
\label{fig:saipe-tunes-vws1-4}
\end{subfigure}
\begin{subfigure}[t]{0.32\textwidth}
\centering
\includegraphics[width=\textwidth]{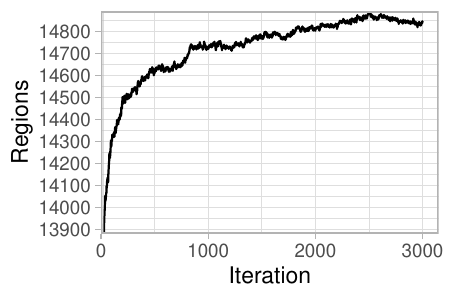}
\caption{}
\label{fig:saipe-comps-vws1-4}
\end{subfigure}

\begin{subfigure}[t]{0.32\textwidth}
\centering
\includegraphics[width=\textwidth]{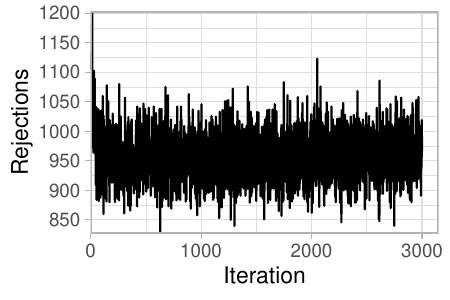}
\caption{}
\label{fig:saipe-rejects-vws1-4}
\end{subfigure}
\begin{subfigure}[t]{0.32\textwidth}
\centering
\includegraphics[width=\textwidth]{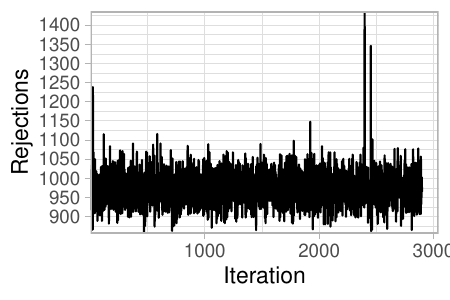}
\caption{}
\label{fig:saipe-rejects-vws2-4-tail}
\end{subfigure}
\end{figure}

A substantially higher computational burden is seen in VWS0 than VWS1 and VWS2, as anticipated. As in Section~\ref{sec:sim-post}, we find that the larger $\epsilon_1$ to permit more rejections tends to improve run time. Additionally, more proactive merging with the larger value of $\epsilon_2$ yields (modest) further improvements. However, a substantial improvement in run time is seen using a limited tuning period. Figure~\ref{fig:saipe-tune} displays the number of tuning steps, regions, and rejections per iteration under VWS1 with $\epsilon_1 = 0.85$ and $\epsilon_2 = 0.01$, each totaled over the $m$ proposals. Additionally, rejections per iteration is shown for VWS2 under the same $\epsilon_1$ and $\epsilon_2$. Figures~\ref{fig:saipe-tune-ext1} and \ref{fig:saipe-tune-ext2} provide a more complete collection of such plots. The number of rejections is seen to not increase substantially after the limited tuning period, aside from jumps in several iterations. With VWS2 and $\epsilon_1 = 0.85$, $\epsilon_2 = 0.01$, the elapsed time with limited tuning is \red{$101.28 / 158.67 \approx 63.8\%$} of when tuning is enabled throughout sampling. This is also \red{$101.28 / 851.56 \approx 11.9\%$} the elapsed time of VWS0. The remainder of this section will utilize VWS results from VWS2 with $\epsilon_1 = 0.85$, $\epsilon_2 = 0.01$.

AMH mixed substantially better in its worst cases than IMH, in terms of both ESS and ESS per second, though both steps can be computed quickly: AMH averaged \red{$\text{30,000} / 40.10 \approx 748.1$} iterations per second while IMH averaged \red{$\text{30,000} / 26.26 \approx \text{1,142.4}$}. The ESS quantiles under AMH are seen to be notably lower than VWS but higher than IMH. Figure~\ref{fig:saipe-ess} shows that the bulk of the ESS values is larger under IMH than AMH, but there are noticeably more elements below \red{250} under IMH than AMH. Trace plots for several elements of $\vec{\sigma}^2$ having the lowest ESS under AMH are displayed in Figure~\ref{fig:saipe-trace-amh}. The speed of AMH iterations places it ahead of VWS0 in ESS per second; however, the faster VWS1 and VWS2 iterations outperform AMH in ESS per second because of their improved mixing.

ARMS is seen to handle this application very well; its elapsed time is \red{$126.55 / 101.28 \approx 124.95\%$} of the fastest VWS run, while mixing of the $\vec{\sigma}^2$ elements is about as good as in VWS. The latter is seen from the distribution of ESS in Figure~\ref{fig:saipe-ess}.

\begin{figure}
\centering
\caption{Estimates and interval widths for $\sigma_1^2, \ldots, \sigma_m^2$ from posterior using VWS versus IMH within-Gibbs. Ratios on vertical axes are based on VWS in the numerator and IMH in the denominator. Horizontal axis is ESS computed under IMH. (\subref{fig:saipe-sigma2-est-imh-vs-vws}) Ratios of posterior means. (\subref{fig:saipe-sigma2-width-imh-vs-vws}) Ratios of credible interval widths. (ESS: effective sample size; VWS: vertical weighted strips; IMH: independent Metropolis-Hastings.)}
\label{fig:saipe-sigma2-imh-vs-vws}
\begin{subfigure}[t]{0.48\textwidth}
\includegraphics[width=\textwidth]{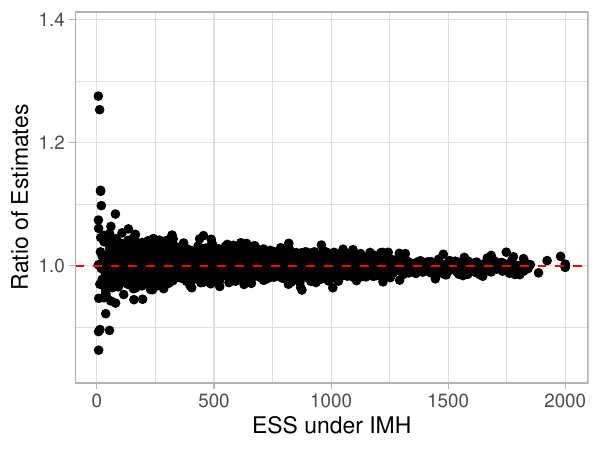}
\caption{}
\label{fig:saipe-sigma2-est-imh-vs-vws}
\end{subfigure}
\begin{subfigure}[t]{0.48\textwidth}
\includegraphics[width=\textwidth]{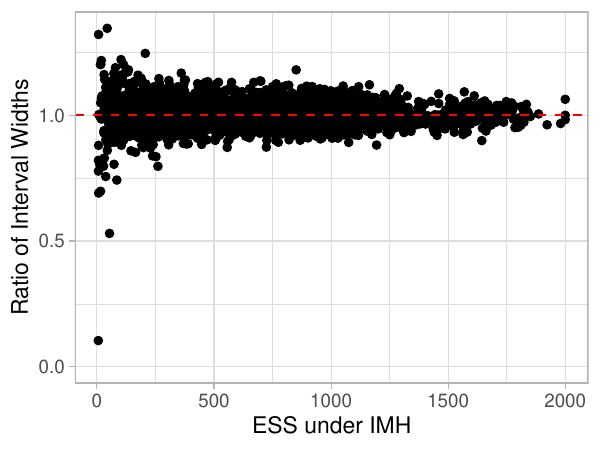}
\caption{}
\label{fig:saipe-sigma2-width-imh-vs-vws}
\end{subfigure}
\end{figure}

The degraded mixing quality of $\vec{\sigma}^2$ under IMH has some effect on summaries of these variables produced using the saved draws. Figure~\ref{fig:saipe-sigma2-imh-vs-vws} compares estimates and credible interval widths of the $\vec{\sigma}^2$ elements from IMH and VWS. To compare estimates, a ratio is computed with the posterior mean under IMH in the numerator and VWS in the denominator. A similar ratio is computed for credible interval widths; here, widths are taken to be the difference between 5\% and 95\% quantiles of $\sigma_i^2$ draws. As we may have anticipated, $\sigma_i^2$ with lower ESS under IMH correspond to ratios with a higher tendency to deviate from 1. Aside from several notable elements with very low ESS, estimates deviate from 1 by a factor of up to about $\pm10\%$ while interval widths deviate by a factor of up to $\pm25\%$.

To summarize, AMH substantially improved the most pathological cases in the IMH analysis and Metropolis steps needed relatively little computation. ARMS required more computation---building a proposal from rejected draws before taking a Metropolis step---but mixed about as well as possible and was readily implemented using the \texttt{armspp} package. Successors to ARMS---such as FUSS with the rejection chain method of proposal construction presented in Table~2 of \citet{Martino-EtAl-2015}---will also mix well. Crucially, VWS provided a means to generate exact draws so that the choice of the within-Gibbs sampling algorithm does not come into question. When tuning via Algorithm~\ref{alg:self-tuned-vws} is kept limited to an initial portion of Gibbs iterations, the computational burden of VWS reduced to below that of ARMS. Therefore, rejection sampling with VWS is useful in this setting not only as a basis for comparison, but also as a practical sampling method.

\begin{table}
\centering
\caption{Comparison of samplers for $\vec{\sigma}^2$ steps. The ``ESS'' column group displays three quantiles of effective sample size over the $m$ chains; ``ESS/sec'' divides these quantities by the elapsed time. Elapsed times are given in seconds. (ESS: effective sample size; VWS: vertical weighted strips; IMH: independent Metropolis-Hastings; ARMS: Adaptive Rejection Metropolis Sampling; AMH: Adaptive Metropolis-Hastings.)}
\label{tab:saipe-mixing}
\begin{tabular}{lrlrrrrrrrr}
\toprule
\multicolumn{3}{c}{} &
\multicolumn{2}{c}{} & 
\multicolumn{3}{c}{ESS $\vec{\sigma}^2$} &
\multicolumn{3}{c}{ESS/sec $\vec{\sigma}^2$} \\
\cmidrule(lr){6-8}
\cmidrule(lr){9-11}
\multicolumn{1}{c}{} &
\multicolumn{1}{c}{$\epsilon_1$} &
\multicolumn{1}{c}{$\epsilon_2$} &
\multicolumn{1}{c}{Rejections} &
\multicolumn{1}{c}{Elapsed} &
\multicolumn{1}{c}{Min} &
\multicolumn{1}{c}{1\%} &
\multicolumn{1}{c}{2.5\%} &
\multicolumn{1}{c}{Min} &
\multicolumn{1}{c}{1\%} &
\multicolumn{1}{c}{2.5\%} \\
\cmidrule(lr){1-1}
\cmidrule(lr){2-3}
\cmidrule(lr){4-5}
\cmidrule(lr){6-8}
\cmidrule(lr){9-11}
IMH  &  --- &   --- & 46,342,848 &  26.26 &     0 &    34 &    73 &  0.0 &  1.3 &  2.8 \\
AMH  &  --- &   --- & 61,926,055 &  40.10 &   190 &   232 &   247 &  4.7 &  5.8 &  6.2 \\
ARMS &  --- &   --- &     12,410 & 126.55 &   796 & 1,641 & 1,728 &  6.3 & 13.0 & 13.7 \\
VWS0 &  --- &   --- &  4,319,988 & 851.56 &   953 & 1,654 & 1,729 &  1.1 &  1.9 &  2.0 \\
\cmidrule(lr){1-1}
\cmidrule(lr){2-3}
\cmidrule(lr){4-5}
\cmidrule(lr){6-8}
\cmidrule(lr){9-11}
VWS1 & 0.75 & 0.001 &  2,248,664 & 174.96 &   949 & 1,655 & 1,735 &  5.4 &  9.5 &  9.9 \\
     & 0.75 & 0.01  &  2,456,020 & 167.96 &   954 & 1,671 & 1,731 &  5.7 &  9.9 & 10.3 \\
     & 0.85 & 0.001 &  2,717,867 & 166.92 & 1,056 & 1,643 & 1,722 &  6.3 &  9.8 & 10.3 \\
     & 0.85 & 0.01  &  2,879,934 & 158.67   & 992 & 1,677 & 1,748 &  6.3 & 10.6 & 11.0 \\
\cmidrule(lr){1-1}
\cmidrule(lr){2-3}
\cmidrule(lr){4-5}
\cmidrule(lr){6-8}
\cmidrule(lr){9-11}
VWS2 & 0.75 & 0.001 &  2,427,008 & 112.61 & 1,037 & 1,632 & 1,727 &  9.2 & 14.5 & 15.3 \\
     & 0.75 & 0.01  &  2,488,942 & 106.06 &   824 & 1,650 & 1,736 &  7.8 & 15.6 & 16.4 \\
     & 0.85 & 0.001 &  2,921,239 & 106.44 &   849 & 1,665 & 1,735 &  8.0 & 15.6 & 16.3 \\
     & 0.85 & 0.01  &  2,928,714 & 101.28 & 1,040 & 1,682 & 1,740 & 10.3 & 16.6 & 17.2 \\
\bottomrule
\end{tabular}
\end{table}

\section{Conclusions}
\label{sec:conclusions}

This paper considered a self-tuned variant of the vertical weighted strips (VWS) method for rejection sampling within Gibbs. By gradually tuning proposal distributions to generate from a collection of evolving conditionals, it can be possible to avoid costly computations of repeatedly constructing and refining new proposals. We found that it is not necessary to seek a small rejection probability with VWS when rejections are relatively inexpensive in relation to tuning; therefore, we only need to ensure that acceptances are not too infrequent. The code used in this work was written in C++ to achieve good performance; however, it is likely that improvements are possible with more careful consideration of data structures and memory use.

We applied self-tuned VWS to a Bayesian model from \citet{You-2021} for small area estimation which utilizes regressions for both the point estimate and sampling variance. The Gibbs sampler from \citet{You-2021} encounters one family of conditionals with an unfamiliar form. Use of the proposed independent Metropolis-Hastings (IMH) step for these conditionals resulted in poor mixing for some of the latent variables in our application. Replacing IMH with exact variate generation using VWS substantially improved mixing. Iterations with self-tuned VWS are more costly than with IMH, but many fewer iterations with exact variate generation are needed to obtain a representative sample for the latent variables. Self-tuned VWS was seen to be substantially faster than VWS with repeated proposal construction: on the order of about \red{1.69} minutes for the former and \red{14.19} minutes for the latter. Adaptive Metropolis-Hastings (AMH) and adaptive rejection Metropolis sampling (ARMS) were also seen to improve mixing in this setting; especially the latter. Exact variate generation can be seen as a best case for mixing within a Gibbs sampler---when it is possible---where we need not be concerned that algorithm choice impairs the chain's convergence to the posterior.

A second application in small area estimation is illustrated in the supplementary materials. Pathological cases of mixing are again encountered in IMH, while ARMS \& AMH do not fare quite as well as in Section~\ref{sec:saipe-analysis}. VWS again facilitates obtaining exact draws within the Gibbs sampler, where variates are drawn exactly from all conditionals. Both applications are similar in that there are $m$ unfamiliar univariate conditionals that correspond to latent variables.

When rejections are relatively cheap, we suggest a larger value for $\epsilon_1$ such as $0.85$ to bound the rejection rate away from 1. Results were less definitive for tuning parameter $\epsilon_2$ used to merge low contribution regions. For a particular dataset, we suggest trying short pilot MCMC runs with several values such as 0.01 and 0.001 to see whether there is a performance difference. An effective strategy was to limit tuning to an initial set of iterations in the MCMC; this reduced overall run times in our applications to \red{55--65\%} of times with tuning used for the entire MCMC. The analyst can monitor acceptance rates to ensure that they do not degrade once tuning is halted. Operations such as majorization for $w$ and integration for $\overline{\xi}_j$ must be done to update proposals between MCMC iterations even when the partition remains unchanged; tuning is more effective as a strategy to reduce run time when these operations are less computationally intensive. It is important to emphasize that $\epsilon_1$, $\epsilon_2$, and the limited tuning period influence run time of the Gibbs sampler but not the quality of draws.

This work utilized a constant majorizer: a VWS proposal which is constructed from piecewise constant functions. One could also consider a self-tuned variant of the linear majorizer which is composed of piecewise linear functions. The linear form often obtains substantially lower rejection rates than the constant form but its implementation is more involved; see \citet*{vws-2026} for details on this construction and several examples where practical proposals are obtained.

This work considered univariate VWS within Gibbs sampling for a posterior distribution of multiple parameters. \citet*{Martino-Read-Luengo-2015} mention the Hit and Run algorithm \citep*{Belisle-Romeijn-Smith-1993} as another multivariate application for adaptable univariate samplers. Here an MCMC chain is generated for a target density $\pi(\vec{x})$ with support $\mathbb{S} \subseteq \mathbb{R}^k$. Relative to the current position $\vec{x}$, a move $\vec{x}' = \vec{x} + \lambda \vec{\theta}$ is determined by sampling direction $\vec{\theta}$ from the unit sphere and length $\lambda \in \mathbb{R}$ from the univariate target $f(\lambda) \propto \pi(\vec{x} + \lambda \vec{\theta}) \ind\{ \vec{x} + \lambda \vec{\theta} \in \mathbb{S} \}$. An $f$ which can be decomposed into a readily sampled base distribution and a majorizable weight function may be an excellent candidate for sampling by VWS within MCMC iterations.

\clearpage
\appendix

\renewcommand{\theequation}{S\arabic{equation}}
\renewcommand{\theexample}{S\arabic{example}}
\renewcommand{\thesection}{S\arabic{section}}
\renewcommand{\thetable}{S\arabic{table}}
\renewcommand{\thefigure}{S\arabic{figure}}
\renewcommand{\thealgorithm}{S\arabic{algorithm}}
\renewcommand{\theremark}{S\arabic{remark}}
\renewcommand{\theproposition}{S\arabic{proposition}}

\setcounter{equation}{0}
\setcounter{example}{0}
\setcounter{section}{0}
\setcounter{table}{0}
\setcounter{figure}{0}
\setcounter{algorithm}{0}
\setcounter{remark}{0}
\setcounter{proposition}{0}

\section{Additional Remarks on VWS}

\begin{figure}
\centering
\caption{Elements of IMH in a case where it is seen to mix poorly. (\subref{fig:imh-density}) Density of $\text{LN}(\mu_i = -1.8177, \tau^2 = 0.0950)$ (solid line) which appears in the acceptance ratio. (\subref{fig:imh-proposal}) CDF of proposal $\text{IG}(\kappa_i = 0.8708, \lambda_i = 1.4072)$. Dashed vertical line represents the 99.9\% quantile $\sigma_i^2 = 0.421$ of $\text{LN}(\mu_i, \tau^2)$. Target \eqref{eqn:target} is shown (dotted line) for reference.}
\label{fig:imh}
\begin{subfigure}[t]{0.40\textwidth}
\includegraphics[width=\textwidth]{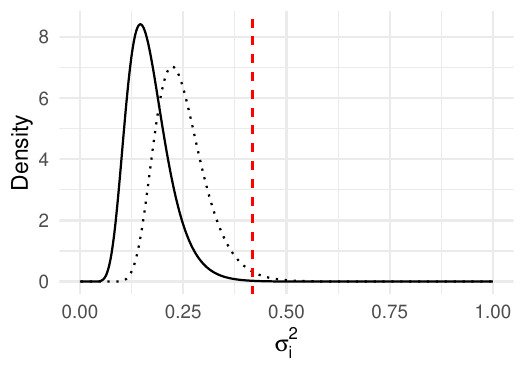}
\caption{}
\label{fig:imh-density}
\end{subfigure}
\begin{subfigure}[t]{0.40\textwidth}
\includegraphics[width=\textwidth]{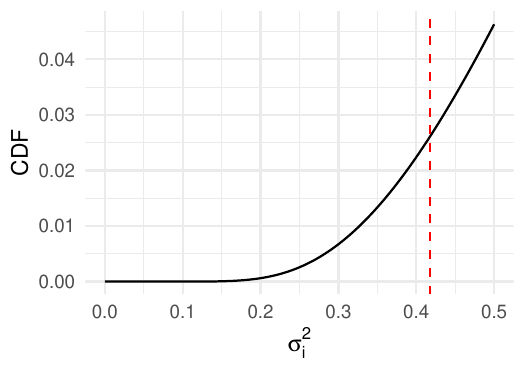}
\caption{}
\label{fig:imh-proposal}
\end{subfigure}
\end{figure}

Figure~\ref{fig:imh} displays the inverse gamma proposal and lognormal density described in Remark~\ref{remark:imh-mix}. The following remarks discuss additional considerations about form \eqref{eqn:target} as a weighted density of these two distributions.

\begin{remark}
\label{remark:decompose}
The choice of decomposition for \eqref{eqn:target} as a weighted density to formulate VWS is not unique. We have utilized a more apparent choice based on the form of the density. It is interesting to mention an alternate choice that leads to a caveat. The factor $(\sigma_i^2)^{-\kappa}$ may be incorporated into the exponential quadratic term of \eqref{eqn:target} as
\begin{align}
f(\sigma_i^2)
&\propto (\sigma_i^2)^{-\kappa_i} \frac{1}{\sigma_i^2}
\exp\{ -\lambda_i / \sigma_i^2 \}
\exp\left\{
-\frac{1}{2 \tau^2} (\log \sigma_i^2 - \mu_i)^2
\right\} \nonumber \\
&\propto \exp\{ -\lambda_i / \sigma_i^2 \}
f_\text{LN}(\sigma_i^2 \mid \mu_i - \kappa_i \tau^2, \tau^2).
\label{eqn:tx-alt}
\end{align}
Here we may consider $w(\sigma_i^2) = e^{-\lambda_i / \sigma_i^2}$ and $T_i \sim \text{LN}(\mu_i - \kappa_i \tau^2, \tau^2)$ as the weight function and base distribution of the weighted density, respectively. When degrees of freedom $d_i$ in the data are large, $\kappa_i = (d_i - 1) / 2$ is also large, and both $\E(T_i) = \exp\{ \mu_i - \kappa_i \tau^2 + \tau^2 / 2 \}$ and $\Var(T_i) = [\exp\{\tau^2\} - 1] \exp\{2(\mu_i - \kappa_i \tau^2) + \tau^2\}$ become extremely small. Numerical issues may be encountered when generating $T_i$ in this case. Therefore, a general precaution when implementing VWS is to check for $g$ which become focused on particular regions of the support; in such cases, an alternative decomposition may be preferred.
\end{remark}

\begin{remark}
Table~\ref{tab:cond-vws} shows that substantially more computation time is needed in the case where $\kappa = 50$ and $\tau = 0.5$. As $\kappa$ is taken larger and $\tau$ is taken smaller, it can be seen that the target distribution \eqref{eqn:target-tx} becomes focused within a small interval near zero which is a low-probability region of base distribution $\text{LN}(\sigma^2 \mid \mu, \tau^2)$. In turn, it requires more draws for the tuning procedure to propose knots that will effectively reduce the rejection rate. One possibility to improve the proposal in this setting is to refactor the target as
\begin{align*}
f(\sigma_i^2) &\propto
f_\text{IG}(\sigma_i^2 \mid \kappa_i, \lambda_i)
f_\text{LN}(\sigma_i^2 \mid \mu_i, \tau^2)
\underbrace{\exp\left\{\frac{1}{\tau^2} \vartheta_i \log \sigma_i^2 \right\}
\exp\left\{-\frac{1}{\tau^2} \vartheta_i \log \sigma_i^2 \right\}}_{=1} \\
&\propto (\sigma_i^2)^{-\kappa_i - 1 - \vartheta_i / \tau^2} \exp\{ -\lambda_i / \sigma_i^2 \}
\frac{1}{\sigma_i^2} \exp\left\{ -\frac{1}{2\tau^2} [(\log \sigma_i^2)^2 - 2 (\mu_i + \vartheta_i) \log \sigma_i^2] \right\} \\
&\propto
\underbrace{f_\text{IG}(\sigma_i^2 \mid \kappa_i + \vartheta_i / \tau^2, \lambda_i)}_{w(\sigma_i^2)}
\underbrace{f_\text{LN}(\sigma_i^2 \mid \mu_i + \vartheta_i, \tau^2)}_{g(\sigma_i^2)}.
\end{align*}
A negative value of the parameter $\vartheta_i \in \mathbb{R}$ can be used to reduce the mass of the base distribution while diminishing the effect of a very large $\kappa_i$ in the weight function. This is not an major issue in the remainder of the work, so we proceed with the proposal based on \eqref{eqn:target-tx}.
\end{remark}

\begin{remark}
Consider the transformation \eqref{eqn:tx-alt} to $\omega_i = -\lambda / \sigma_i^2$ which yields
\begin{align}
f(\omega_i^2)
&\propto \exp\{ -\omega_i^2 \}
f_\text{LN}(\omega_i^2 \mid \log \lambda_i - \mu_i + \kappa_i \tau^2, \tau^2),
\label{eqn:tx-alt2}
\end{align}
using Jacobian $\partial \sigma_i^2 / \partial \omega_i^2 = -\lambda_i / (\omega_i^2)^2$. We recognize \eqref{eqn:tx-alt2} as the integrand of a lognormal moment-generating function. Approximation of such integrals is a topic of interest because they appear not to have a closed form \citep*[e.g.,][]{Asmussen-EtAl-2016}. Because \eqref{eqn:tx-alt2} is a transformation of \eqref{eqn:target-tx}, it appears that a numerical method of integration is necessary to compute normalizing constant $\psi$ for \eqref{eqn:target-tx}. However, $\psi$ is not needed in VWS or the self-tuned variant; in particular, bound $\rho_+$ may be used in place of $\psi$ to monitor probability of rejection.
\end{remark}

\clearpage

\section{Conditional Simulation: Additional Results}
\label{sec:sim-cond-extra}

\begin{figure}
\centering
\caption{Results of first $n = 20$ draws using self-tuned VWS, plotted as a median over 10,000 repetitions. Left subfigures display log-bound with nominal level $\epsilon_1$ as a dashed line and right subfigures show corresponding number of regions in the proposal. Symbols correspond to values of $(\kappa, \tau)$: ``$\Circle$'' is $(\kappa = 10, \tau = 0.5)$, ``$\triangle$'' is $(\kappa = 10, \tau = 1.0)$, ``$+$'' is $(\kappa = 50, \tau = 0.5)$, and ``$\times$'' is $(\kappa = 50, \tau = 1.0)$.}
\label{fig:cond-vws}
\begin{subfigure}[t]{\textwidth}
\centering
\includegraphics[width=0.38\textwidth]{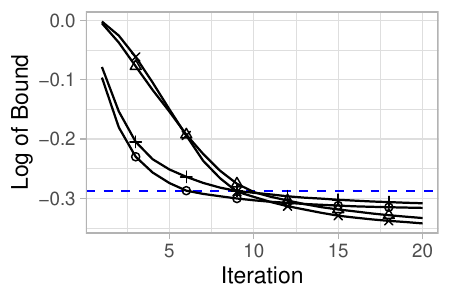}
\includegraphics[width=0.38\textwidth]{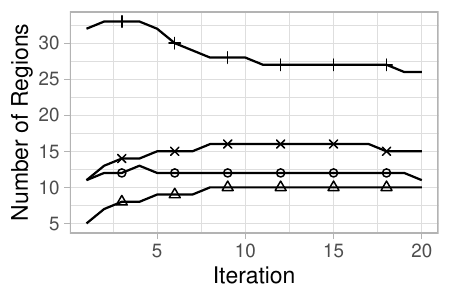}
\caption{$\epsilon_1 = 0.75, \epsilon_2 = 0.01.$}
\label{fig:cond-vws-1-1}
\end{subfigure}
\begin{subfigure}[t]{\textwidth}
\centering
\includegraphics[width=0.38\textwidth]{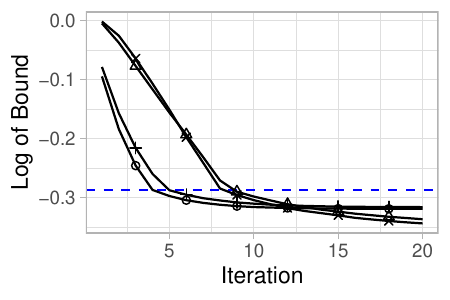}
\includegraphics[width=0.38\textwidth]{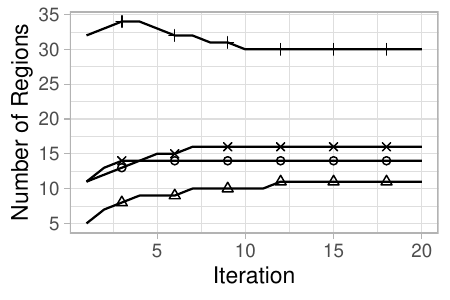}
\caption{$\epsilon_1 = 0.75, \epsilon_2 = 0.001.$}
\label{fig:cond-vws-1-2}
\end{subfigure}
\begin{subfigure}[t]{\textwidth}
\centering
\includegraphics[width=0.38\textwidth]{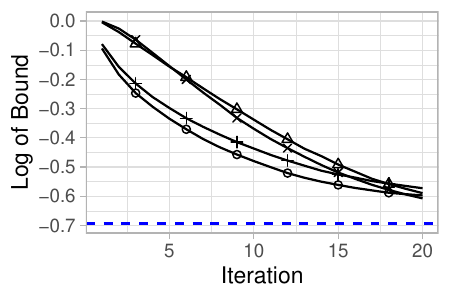}
\includegraphics[width=0.38\textwidth]{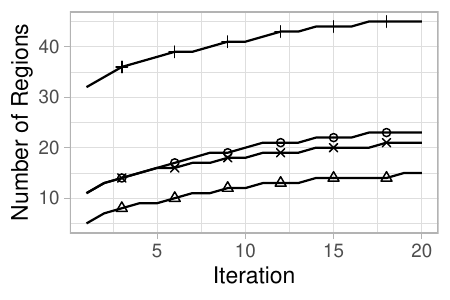}
\caption{$\epsilon_1 = 0.50, \epsilon_2 = 0.01.$}
\label{fig:cond-vws-2-1}
\end{subfigure}
\begin{subfigure}[t]{\textwidth}
\centering
\includegraphics[width=0.38\textwidth]{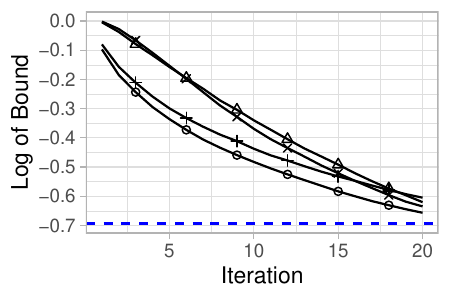}
\includegraphics[width=0.38\textwidth]{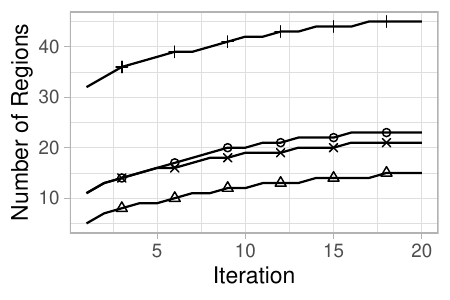}
\caption{$\epsilon_1 = 0.50, \epsilon_2 = 0.001.$}
\label{fig:cond-vws-2-2}
\end{subfigure}
\end{figure}

Figure~\ref{fig:cond-vws} summarizes the log of bound \eqref{eqn:bound} and the number of regions averaged over the $R$ chains generated in Section~\ref{sec:sim-cond}.

\section{Posterior Simulation: Additional Results}
\label{sec:sim-post-extra}

Table~\ref{tab:sim-post-ext} provides some additional summaries for draws of the posterior simulation presented in Section~\ref{sec:sim-post}. Summaries include the ESS for draws of the mean $\vec{\vartheta}$, ESS per second for draws of $\sigma^2$ corresponding to ESS in Table~\ref{tab:sim-post}, and multivariate ESS (mESS) for $\vec{\theta}$. As expected, VWS chains are on par or higher in ESS than IMH for both $\vec{\vartheta}$ and $\vec{\theta}$ for all $\epsilon_1$, $\epsilon_2$, and $m$. For the three displayed quantiles of ESS per second for $\vec{\sigma^2}$, VWS obtains higher values for both $m = 500$ and $m = \text{2,000}$. The best values are obtained with $\epsilon_1$ taken to the largest setting $0.85$.

\begin{table}
\centering
\caption{Additional results from posterior simulation. Metrics are averaged over $S = 500$ simulation repetitions. The ``ESS $\vec{\vartheta}$'' column group represents three quantiles of the univariate ESS metric of $\vec{\vartheta}$ entries and have been rounded to whole numbers. Column ``mESS $\vec{\theta}$'' is the multivariate ESS for parameter $\vec{\theta}$.}
\label{tab:sim-post-ext}
\begin{tabular}{rrlrrrrrrr}
\toprule
\multicolumn{3}{l}{(a) IMH} &
\multicolumn{3}{c}{ESS $\vec{\vartheta}$} &
\multicolumn{3}{c}{ESS/sec $\vec{\sigma}^2$} &
\multicolumn{1}{c}{} \\
\cmidrule(lr){4-6}
\cmidrule(lr){7-9}
\multicolumn{1}{c}{$m$} &
\multicolumn{1}{c}{} &
\multicolumn{1}{c}{} &
\multicolumn{1}{c}{Min} &
\multicolumn{1}{c}{1\%} &
\multicolumn{1}{c}{2.5\%} &
\multicolumn{1}{c}{Min} &
\multicolumn{1}{c}{1\%} &
\multicolumn{1}{c}{2.5\%} &
\multicolumn{1}{c}{mESS $\vec{\theta}$} \\
\cmidrule(lr){1-3}
\cmidrule(lr){4-6}
\cmidrule(lr){7-9}
\cmidrule(lr){10-10}
  500 &  &  & 644 & 1,001 & 1,298 & 4.6 & 10.5 & 15.5 & 254 \\
2,000 &  &  & 971 & 1,583 & 1,693 & 0.7 &  2.7 &  4.1 & 254 \\
\midrule
\multicolumn{3}{l}{(b) VWS} &
\multicolumn{3}{c}{ESS $\vec{\vartheta}$} &
\multicolumn{3}{c}{ESS/sec $\vec{\sigma}^2$} &
\multicolumn{1}{c}{} \\
\cmidrule(lr){4-6}
\cmidrule(lr){7-9}
\multicolumn{1}{c}{$m$} &
\multicolumn{1}{c}{$\epsilon_1$} &
\multicolumn{1}{c}{$\epsilon_2$} &
\multicolumn{1}{c}{Min} &
\multicolumn{1}{c}{1\%} &
\multicolumn{1}{c}{2.5\%} &
\multicolumn{1}{c}{Min} &
\multicolumn{1}{c}{1\%} &
\multicolumn{1}{c}{2.5\%} &
\multicolumn{1}{c}{mESS $\vec{\theta}$} \\
\cmidrule(lr){1-3}
\cmidrule(lr){4-6}
\cmidrule(lr){7-9}
\cmidrule(lr){10-10}
  500 & 0.50 & 0.0001 &   677 & 1,031 & 1,322 & 46.9 & 53.6 & 55.5 & 388 \\
  500 & 0.50 & 0.001  &   698 & 1,053 & 1,330 & 50.5 & 56.9 & 58.9 & 388 \\
  500 & 0.50 & 0.01   &   685 & 1,036 & 1,335 & 46.9 & 53.8 & 55.7 & 384 \\
  500 & 0.75 & 0.0001 &   675 & 1,043 & 1,329 & 64.1 & 72.9 & 75.5 & 389 \\
  500 & 0.75 & 0.001  &   682 & 1,038 & 1,319 & 65.3 & 73.1 & 75.7 & 382 \\
  500 & 0.75 & 0.01   &   694 & 1,054 & 1,339 & 63.2 & 71.7 & 74.2 & 380 \\
  500 & 0.85 & 0.0001 &   668 & 1,022 & 1,304 & 69.7 & 79.1 & 81.9 & 378 \\
  500 & 0.85 & 0.001  &   672 & 1,035 & 1,318 & 69.0 & 78.6 & 81.5 & 382 \\
  500 & 0.85 & 0.01   &   672 & 1,025 & 1,304 & 69.7 & 79.0 & 81.9 & 383 \\
2,000 & 0.50 & 0.0001 & 1,066 & 1,620 & 1,715 & 11.4 & 13.6 & 14.0 & 399 \\
2,000 & 0.50 & 0.001  & 1,051 & 1,619 & 1,714 & 11.6 & 14.0 & 14.5 & 389 \\
2,000 & 0.50 & 0.01   & 1,063 & 1,620 & 1,715 & 11.4 & 13.7 & 14.1 & 389 \\
2,000 & 0.75 & 0.0001 & 1,061 & 1,623 & 1,718 & 15.0 & 17.9 & 18.5 & 395 \\
2,000 & 0.75 & 0.001  & 1,059 & 1,620 & 1,716 & 15.1 & 17.8 & 18.4 & 395 \\
2,000 & 0.75 & 0.01   & 1,057 & 1,621 & 1,716 & 15.0 & 17.9 & 18.5 & 391 \\
2,000 & 0.85 & 0.0001 & 1,045 & 1,621 & 1,715 & 16.3 & 19.4 & 20.1 & 384 \\
2,000 & 0.85 & 0.001  & 1,051 & 1,621 & 1,716 & 16.1 & 19.1 & 19.8 & 389 \\
2,000 & 0.85 & 0.01   & 1,064 & 1,621 & 1,715 & 16.2 & 19.2 & 19.9 & 392 \\
\bottomrule
\end{tabular}
\end{table}

\clearpage

\section{SAIPE Application: Additional Results}
\label{sec:saipe-extra}
This section includes some additional results and summaries from the SAIPE application discussed in Section~\ref{sec:saipe-analysis}.

Table~\ref{tab:saipe-theta-ess} summarizes mixing for parameters $\vec{\theta}$ across the different samplers. ESS for $\phi^2$ and elements of $\vec{\beta}$ show relatively good mixing across all samplers. For $\vec{\gamma}$, ESS is somewhat lower under IMH and AMH than other options. However, ESS for $\tau^2$ is quite variable among the samplers. Therefore, mixing of $\gamma_1$ and $\gamma_2$ may have been affected by the presence of slow-mixing $\vec{\sigma}^2$ elements, but it is not clear if this was also the case for $\tau^2$.

\begin{table}
\centering
\caption{Effective sample size of parameters in $\vec{\theta}$ under varying samplers for the $\vec{\sigma}^2$ steps.}
\label{tab:saipe-theta-ess}
\begin{tabular}{lrlrrrrrrr}
\toprule
\multicolumn{1}{c}{} &
\multicolumn{1}{c}{$\epsilon_1$} &
\multicolumn{1}{c}{$\epsilon_2$} &
\multicolumn{1}{c}{$\beta_1$} &
\multicolumn{1}{c}{$\beta_2$} &
\multicolumn{1}{c}{$\beta_3$} &
\multicolumn{1}{c}{$\gamma_1$} &
\multicolumn{1}{c}{$\gamma_2$} &
\multicolumn{1}{c}{$\phi^2$} &
\multicolumn{1}{c}{$\tau^2$} \\
\cmidrule(lr){1-1}
\cmidrule(lr){2-3}
\cmidrule(lr){4-10}
IMH  &  --- &   --- & 1,313 &   987 & 1,008 & 157 & 183 & 633 & 111 \\
AMH  &  --- &   --- &   981 &   976 &   924 & 134 & 151 & 653 & 130 \\
ARMS &  --- &   --- &   962 &   999 & 1,000 & 279 & 348 & 698 & 257 \\
VWS0 &  --- &   --- & 1,031 & 1,198 & 1,462 & 303 & 327 & 715 & 221 \\
\cmidrule(lr){1-1}
\cmidrule(lr){2-3}
\cmidrule(lr){4-10}
VWS1 & 0.75 & 0.001 &   871 &   919 & 1,104 & 300 & 429 & 721 & 142 \\
     & 0.75 & 0.01  & 1,194 & 1,118 & 1,166 & 307 & 285 & 831 & 127 \\
     & 0.85 & 0.001 & 1,035 &   978 & 1,100 & 320 & 386 & 644 & 183 \\
     & 0.85 & 0.01  &   901 &   787 &   926 & 270 & 316 & 830 & 199 \\
\cmidrule(lr){1-1}
\cmidrule(lr){2-3}
\cmidrule(lr){4-10}
VWS2 & 0.75 & 0.001 & 1,037 & 1,041 &   945 & 288 & 313 & 750 & 144 \\
     & 0.75 & 0.01  &   917 &   980 &   940 & 270 & 326 & 802 & 187 \\
     & 0.85 & 0.001 &   955 &   967 &   892 & 393 & 367 & 677 & 122 \\
     & 0.85 & 0.01  & 1,055 & 1,036 & 1,054 & 353 & 417 & 605 & 468 \\
\bottomrule
\end{tabular}
\end{table}

\begin{figure}
\centering
\caption{Rejections per Gibbs iteration using VWS0.}
\label{fig:saipe-tune-ext0}
\includegraphics[width=0.4\textwidth]{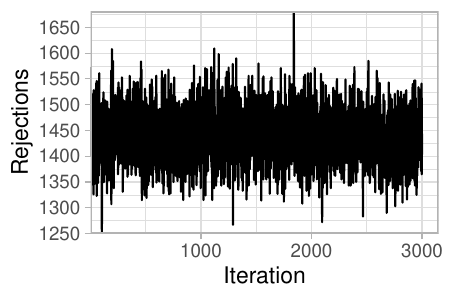}
\end{figure}

\begin{figure}
\centering
\caption{Several metrics by Gibbs iteration using VWS1.
Columns correspond to metrics:
(1) number of tuning adjustments,
(2) number of regions, and
(3) number of rejections.
Rows correspond to tuning parameter settings: 
(1) $\epsilon_1 = 0.75$, $\epsilon_2 = 0.001$;
(2) $\epsilon_1 = 0.75$, $\epsilon_2 = 0.01$;
(3) $\epsilon_1 = 0.85$, $\epsilon_2 = 0.001$;
(4) $\epsilon_1 = 0.85$, $\epsilon_2 = 0.01$.
Initial iterations are omitted from each plot to aid visualization.
}
\label{fig:saipe-tune-ext1}
\begin{tabular}{p{1mm}ccc}
& \textbf{1} & \textbf{2} & \textbf{3} \\
\vspace {-0.85in} \textbf{1} &
\includegraphics[width=0.32\textwidth]{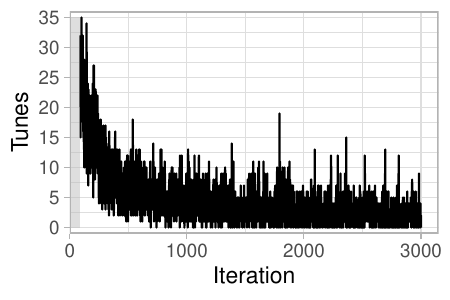} &
\includegraphics[width=0.32\textwidth]{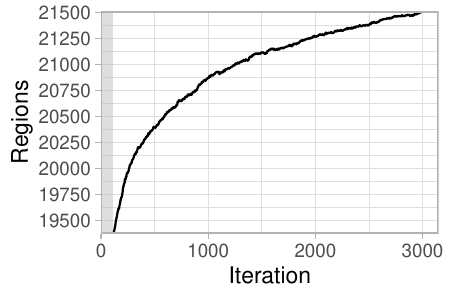} &
\includegraphics[width=0.32\textwidth]{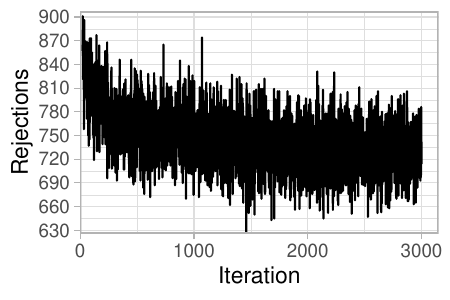} \\
\vspace {-0.85in} \textbf{2} &
\includegraphics[width=0.32\textwidth]{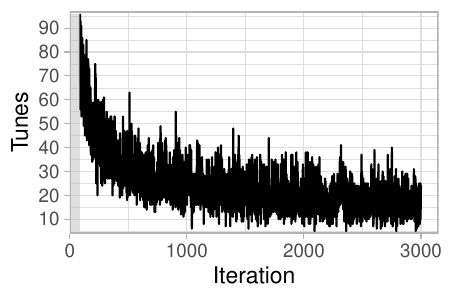} &
\includegraphics[width=0.32\textwidth]{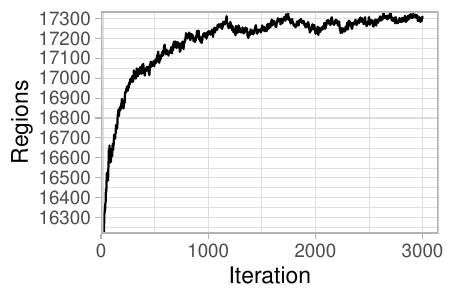} &
\includegraphics[width=0.32\textwidth]{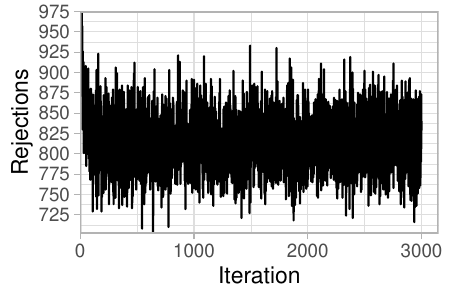} \\
\vspace {-0.85in} \textbf{3} &
\includegraphics[width=0.32\textwidth]{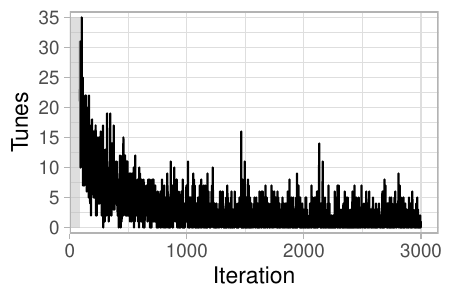} &
\includegraphics[width=0.32\textwidth]{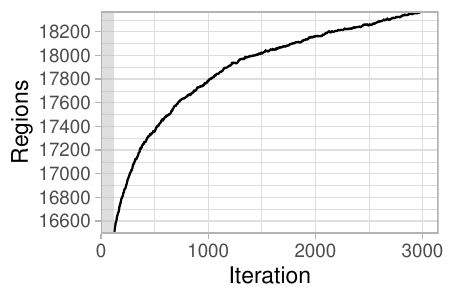} &
\includegraphics[width=0.32\textwidth]{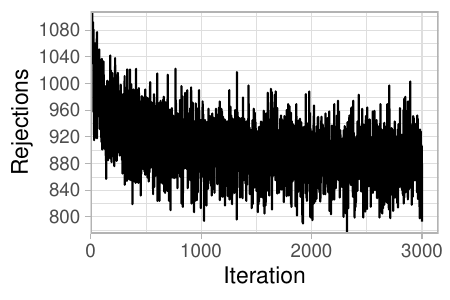} \\
\vspace {-0.85in} \textbf{4} &
\includegraphics[width=0.32\textwidth]{figs4j.pdf} &
\includegraphics[width=0.32\textwidth]{figs4k.pdf} &
\includegraphics[width=0.32\textwidth]{figs4l.pdf} \\
\end{tabular}
\end{figure}

Figure~\ref{fig:saipe-tune-ext0} plots the number of rejections per iteration for VWS0. The number is somewhat higher than in the self-tuning variants. The inclusion of self-tuning in VWS1 has a comparatively lower number of rejections, which also decreases as the MCMC chain evolves, as shown in the third column of Figure~\ref{fig:saipe-tune-ext1}. This figure also includes summaries of the number of tuning adjustments as well as the number of regions by iteration for the four different tuning parameter settings. 

\begin{figure}
\centering
\caption{Rejections per Gibbs iteration using VWS2. Initial iterations are omitted from each plot to aid visualization.}
\label{fig:saipe-tune-ext2}
\begin{subfigure}[t]{0.32\textwidth}
\includegraphics[width=\textwidth]{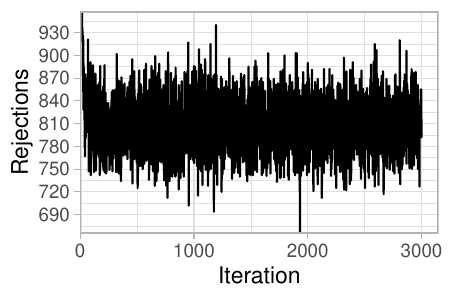}
\caption{$\epsilon_1 = 0.75$, $\epsilon_2 = 0.001$.}
\end{subfigure}
\begin{subfigure}[t]{0.32\textwidth}
\includegraphics[width=\textwidth]{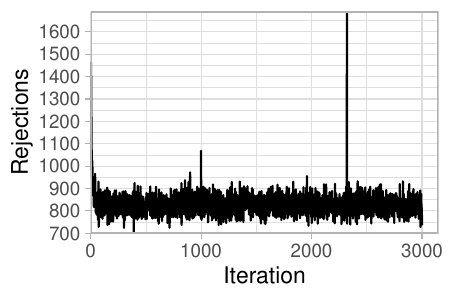}
\caption{$\epsilon_1 = 0.75$, $\epsilon_2 = 0.01$.}
\end{subfigure}

\begin{subfigure}[t]{0.32\textwidth}
\includegraphics[width=\textwidth]{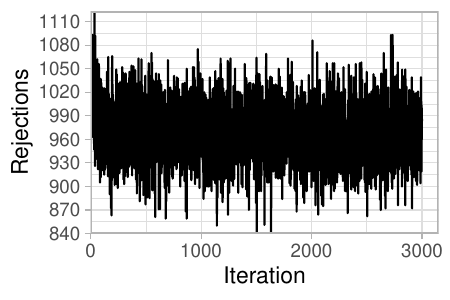}
\caption{$\epsilon_1 = 0.85$, $\epsilon_2 = 0.001$.}
\end{subfigure}
\begin{subfigure}[t]{0.32\textwidth}
\includegraphics[width=\textwidth]{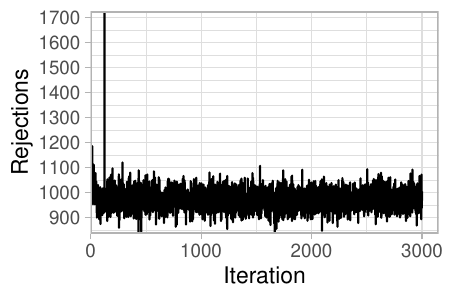}
\caption{$\epsilon_1 = 0.85$, $\epsilon_2 = 0.01$.}
\end{subfigure}
\end{figure}

Figure~\ref{fig:saipe-tune-ext2} shows that limiting the number of tuning steps does not appear to greatly affect the rejection rate for any of the four tuning parameter settings compared to the rejection plots in Figure~\ref{fig:saipe-tune-ext1}. In particular, for the two cases when $\epsilon_2 = 0.001$, there are no noticeable spikes or upward trends in rejections per iteration, and only several spikes when $\epsilon_2 = 0.01$.

\begin{figure}
\centering
\caption{The three chains of $\vec{\sigma}^2$ with lowest ESS under AMH.}
\label{fig:saipe-trace-amh}
\includegraphics[width=0.32\textwidth]{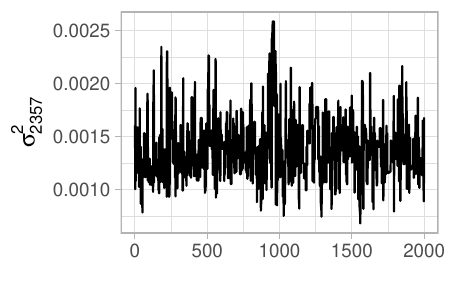}
\includegraphics[width=0.32\textwidth]{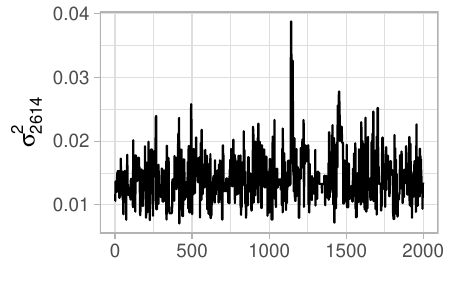}
\includegraphics[width=0.32\textwidth]{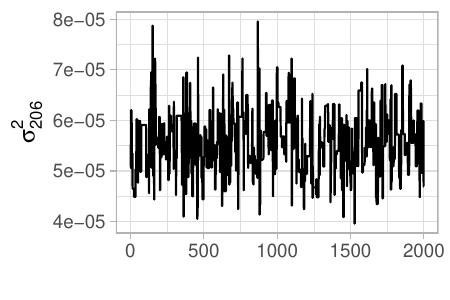}
\end{figure}

\begin{figure}
\centering
\caption{The three chains of $\vec{\sigma}^2$ with lowest ESS under ARMS.}
\label{fig:saipe-trace-arms}
\includegraphics[width=0.32\textwidth]{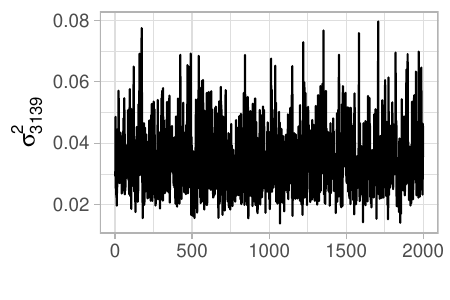}
\includegraphics[width=0.32\textwidth]{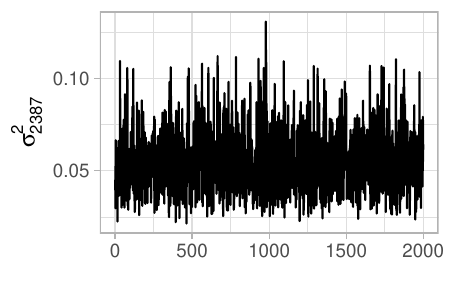}
\includegraphics[width=0.32\textwidth]{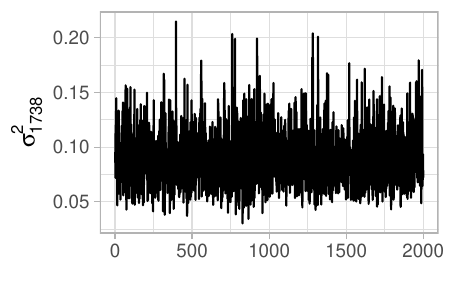}
\end{figure}

Figures~\ref{fig:saipe-trace-amh} and \ref{fig:saipe-trace-arms} display the three chains for $\vec{\sigma}^2$ with the lowest ESS under AMH and ARMS, respectively. The chains from AMH are substantially improved from IMH but can be seen to mixing more slow than those in ARMS and VWS.

\begin{table}
\centering
\caption{Posterior summary of saved draws for parameters in $\vec{\theta}$ using VWS1 with $\epsilon_1 = 0.85$ and $\epsilon_2 = 0.01$. The columns ``SD'', ``5\%'' and ``95\%'' represent the standard deviation, 5\% quantile, and 95\% quantile, respectively.}
\label{tab:saipe-summary}
\begin{tabular}{rrrrr}
\toprule
\multicolumn{1}{c}{} &
\multicolumn{1}{c}{Mean} &
\multicolumn{1}{c}{SD} &
\multicolumn{1}{c}{5\%} &
\multicolumn{1}{c}{95\%} \\
\cmidrule(lr){1-1}
\cmidrule(lr){2-5}
$\beta_1$  & -0.3976 & 0.0387 & -0.4632 & -0.3330 \\
$\beta_2$  &  0.5135 & 0.0078 &  0.5004 &  0.5259 \\
$\beta_3$  &  0.4344 & 0.0086 &  0.4206 &  0.4485 \\
$\gamma_1$ &  1.8439 & 0.0560 &  1.7526 &  1.9344 \\
$\gamma_2$ & -0.9024 & 0.0073 & -0.9142 & -0.8904 \\
$\phi^2$   &  0.0516 & 0.0017 &  0.0489 &  0.0544 \\
$\tau^2$   &  0.0953 & 0.0054 &  0.0865 &  0.1047 \\
\bottomrule
\end{tabular}
\end{table}

\clearpage

\section{Unmatched SAE Model}
\label{sec:unmatch}

Self-tuned VWS may be well-suited to applications beyond the joint SAE model presented previously which involve draws from non-standard conditional distributions. A second such illustration is based on a model presented by \citet{You-Rao-2002}. Here the sampling and linking models for a Fay-Heriot model are ``unmatched'' in that they utilize non-conjugate components beyond the usual linear mixed model setting. This work applies the model to a study of 1991 Canadian census undercoverage data for the 1990 Canadian census. Here the true (latent) undercoverage count is assumed to be measured with error in the sampling model. However, the linking model is formulated as a model for the undercoverage rate, rather than as a count which necessitates the use of an unmatched approach.

Section~\ref{sec:unmatch-model} presents one particular model from this setting and a Gibbs sampler with one non-standard family of conditionals. The IMH, AMH, ARMS, and VWS methods compared under the joint SAE model are again considered in Section~\ref{sec:unmatch-illustration}.

\subsection{Model and Gibbs Sampler}
\label{sec:unmatch-model}

The sampling and linking models they proposed are respectively given by
\begin{align*}
&y_i = \mu_i + \epsilon_i, \quad \epsilon_i \sim \text{N}(0, \sigma_i^2), \\
&k(\mu_i) = \vec{x}_i^\top \vec{\beta} + v_i, \quad v_i \sim \text{N}(0, \tau^2),
\end{align*}
where $\sigma_i^2$ are known fixed numbers and $\vec{x}_i \in \mathbb{R}^d$, with the flat prior $\pi(\vec{\beta}) \propto 1$ and $\tau^2 \sim \text{IG}(a, b)$ and where the link function $g$ is known. For this illustration, we assume the log-link $k(x) = \log x$. The joint distribution of all random variables is
\begin{align*}
f(\vec{y}, \vec{\mu}, \vec{\beta}, \tau^2) \propto
(\tau^2)^{-a-1} e^{-b/\tau^2}
\prod_{i=1}^m \frac{1}{\sigma_i \sqrt{2\pi}}
\exp\left\{
-\frac{1}{2 \sigma_i^2} (y_i - \mu_i)^2
\right\}
\frac{1}{\mu_i \tau \sqrt{2\pi}}
\exp\left\{
-\frac{1}{2 \tau^2} \{ \log \mu_i - \vec{x}_i^\top \vec{\beta} \}^2
\right\}.
\end{align*}
From here, we obtain conditional distributions:
\begin{enumerate}
\item $[\tau^2 \mid \text{---}] \sim \text{IG}(a', b')$ where $a' = a + \frac{m}{2}$,
\begin{math}
b' = b + \frac{1}{2} \{\log \vec{\mu} - \vec{X} \vec{\beta}\}^\top
\{\log \vec{\mu} - \vec{X} \vec{\beta}\},
\end{math}
and $\log \vec{\mu} = [\log \mu_1, \ldots, \log \mu_m]$. 

\item $[\vec{\beta} \mid \text{---}] \sim \text{N}(\vec{\vartheta}, \vec{\Omega}^{-1})$ where
$\vec{\Omega} = \tau^{-2} \vec{X}^\top \vec{X}$ and
$\vec{\vartheta} = (\vec{X}^\top \vec{X})^{-1} \vec{X}^\top ( \log \vec{\mu} )$.

\item $[\vec{\mu} \mid \text{---}] \sim \prod_{i=1}^m [\mu_i \mid \text{---}]$ where
\begin{align}
f(\mu_i \mid \text{---}) \propto
\frac{1}{\sigma_i \sqrt{2\pi}}
\exp\left\{
-\frac{1}{2 \sigma_i^2} (y_i - \mu_i)^2
\right\}
\frac{1}{\mu_i \tau \sqrt{2\pi}}
\exp\left\{
-\frac{1}{2 \tau^2} \{ \log \mu_i - \vec{x}_i^\top \vec{\beta} \}^2
\right\}.
\label{eqn:unmatch-target}
\end{align}
\end{enumerate}
The conditional \eqref{eqn:unmatch-target} does not appear to be a familiar distribution with a well-established variate generation procedure. \citet{You-Rao-2002} consider an IMH algorithm with proposal distribution $\mu_i^* \sim \text{N}(y_i, \sigma_i^2)$ and acceptance probability
\begin{align*}
\varpi_i^{(r)}
&= \min\left\{1, \frac{
f_\text{LN}(\mu_i^* \mid \vec{x}_i^\top \vec{\beta}, \tau^2)
}{
f_\text{LN}(\mu_i^{(r)} \mid \vec{x}_i^\top \vec{\beta}, \tau^2)
} \right\}.
\end{align*}
For AMH, we draw proposal variate $\mu_i^* \sim \text{N}(\mu^{(r)}, v_i^{(r)})$ and accept with probability
\begin{align*}
\varpi_i^{(r)}
&= \min\left\{ 1, \frac{
f_\text{LN}(\mu_i^* \mid \vec{x}_i^\top \vec{\beta}, \tau^2) \cdot
f_\text{N}(\mu_i^* \mid y_i, \sigma_i^2)
}{
f_\text{LN}(\mu^{(r)} \mid \vec{x}_i^\top \vec{\beta}, \tau^2) \cdot
f_\text{N}(\mu^{(r)} \mid y_i, \sigma_i^2)
} \right\}.
\end{align*}
For VWS, we use the decomposition $f(\mu_i \mid \text{---}) \propto w(\mu_i) g(\mu_i)$ where $g(\mu_i) = \text{N}(\mu_i \mid y_i, \sigma_i^2)$ and $w(\mu_i) = \text{LN}(\mu_i \mid \vec{x}_i^\top \vec{\beta}, \tau^2)$, and proceed as in Section~\ref{sec:self-tune-vws}.

\subsection{Illustration}
\label{sec:unmatch-illustration}

We now illustrate the performance of several within-Gibbs samplers for the non-conjugate distribution \eqref{eqn:unmatch-target} using a dataset generated from the model in Section~\ref{sec:unmatch-model}. Suppose there are $m = \text{2,000}$ small areas with sampling variances $\sigma_i^2 \iid \text{Gamma}(1.25, 1/20)$, with expected value is $1.25 \cdot 20 = 25$. The covariate is taken to be $\vec{X} = (\vec{1}, \vec{x}_2)$ where $\vec{x}_2 \iid \text{N}(0, 1)$. Data-generating parameters are $\vec{\beta} = (1, -1)$ and $\tau^2 = 1.25^2$.

\begin{figure}
\centering
\caption{Empirical CDF of ESS for $\vec{\mu}$: IMH ($\CIRCLE$), AMH ($\blacktriangle$), ARMS ($\blacklozenge$) and VWS ($\blacksquare$).}
\label{fig:unmatch-ess}
\includegraphics[width=0.5\textwidth]{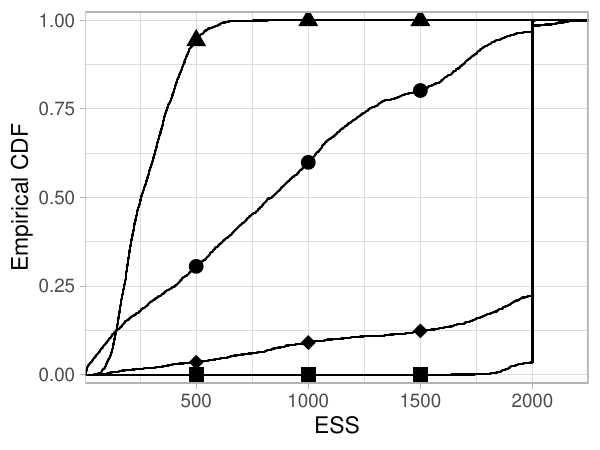}
\end{figure}

\begin{figure}
\centering
\caption{Rejections per Gibbs iteration using VWS0.}
\label{fig:unmatch-tune-ext0}
\includegraphics[width=0.4\textwidth]{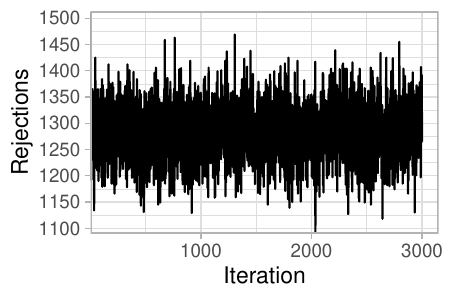}
\end{figure}

\begin{figure}
\centering
\caption{Several metrics by Gibbs iteration using VWS1.
Columns correspond to metrics:
(1) number of tuning adjustments,
(2) number of regions, and
(3) number of rejections.
Rows correspond to tuning parameter settings: 
(1) $\epsilon_1 = 0.75$, $\epsilon_2 = 0.001$;
(2) $\epsilon_1 = 0.75$, $\epsilon_2 = 0.01$;
(3) $\epsilon_1 = 0.85$, $\epsilon_2 = 0.001$;
(4) $\epsilon_1 = 0.85$, $\epsilon_2 = 0.01$.
Initial iterations are omitted from each plot to aid visualization.}
\label{fig:unmatch-tune-ext1}
\begin{tabular}{p{1mm}ccc}
& \textbf{1} & \textbf{2} & \textbf{3} \\
\vspace {-0.85in} \textbf{1} &
\includegraphics[width=0.32\textwidth]{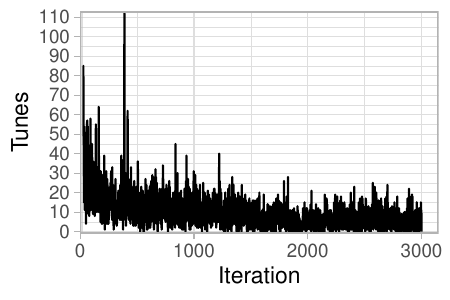} &
\includegraphics[width=0.32\textwidth]{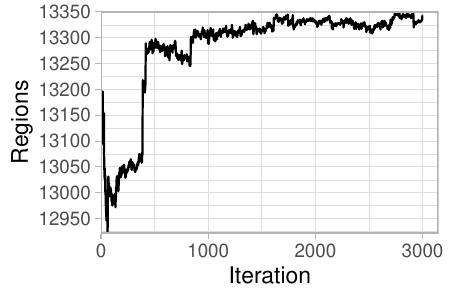} &
\includegraphics[width=0.32\textwidth]{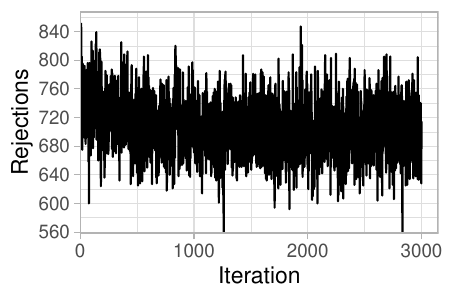} \\
\vspace {-0.85in} \textbf{2} &
\includegraphics[width=0.32\textwidth]{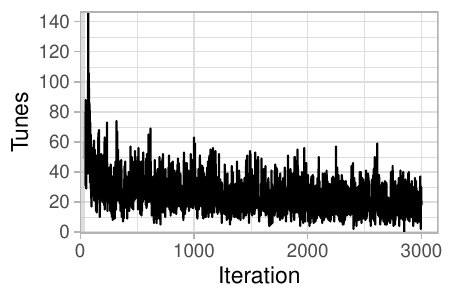} &
\includegraphics[width=0.32\textwidth]{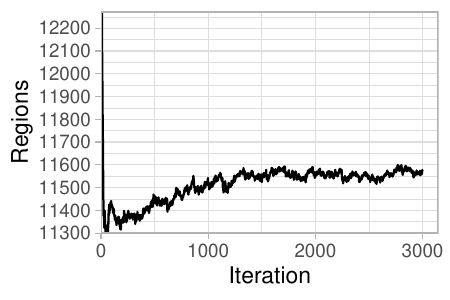} &
\includegraphics[width=0.32\textwidth]{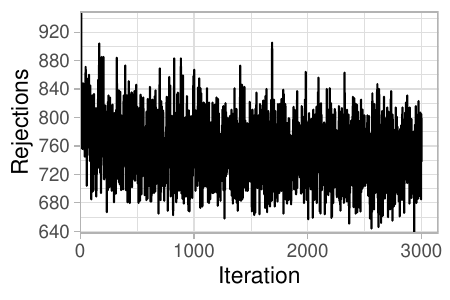} \\
\vspace {-0.85in} \textbf{3} &
\includegraphics[width=0.32\textwidth]{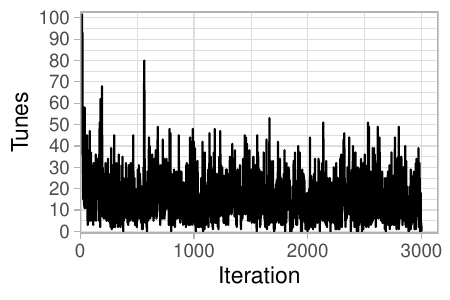} &
\includegraphics[width=0.32\textwidth]{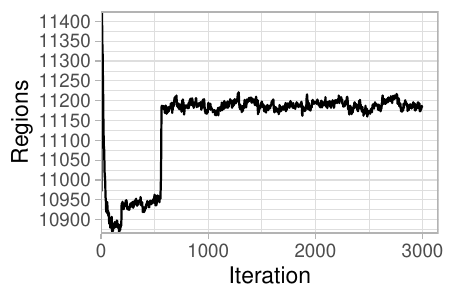} &
\includegraphics[width=0.32\textwidth]{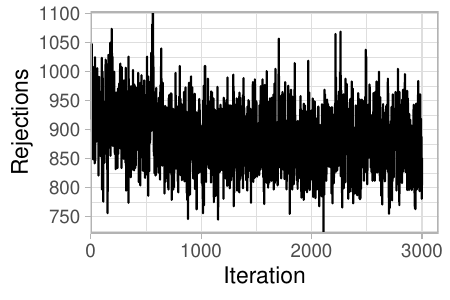} \\
\vspace {-0.85in} \textbf{4} &
\includegraphics[width=0.32\textwidth]{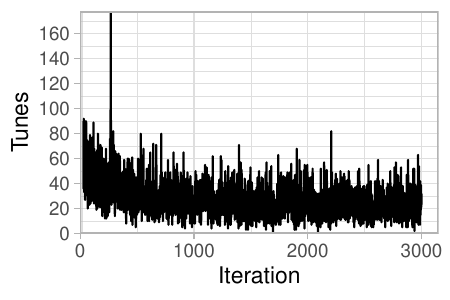} &
\includegraphics[width=0.32\textwidth]{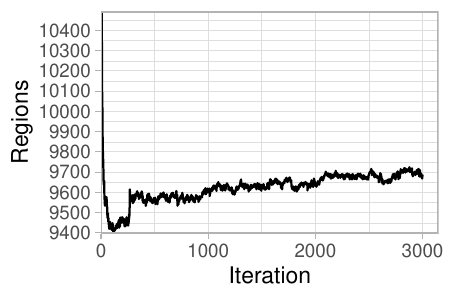} &
\includegraphics[width=0.32\textwidth]{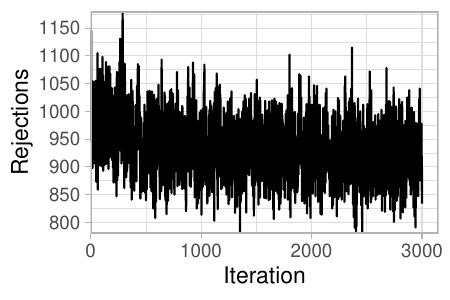} \\
\end{tabular}
\end{figure}

\begin{figure}
\centering
\caption{Rejections per Gibbs iteration using VWS2. Initial iterations are omitted from each plot to aid visualization.}
\label{fig:unmatch-tune-ext2}
\begin{subfigure}[t]{0.32\textwidth}
\includegraphics[width=\textwidth]{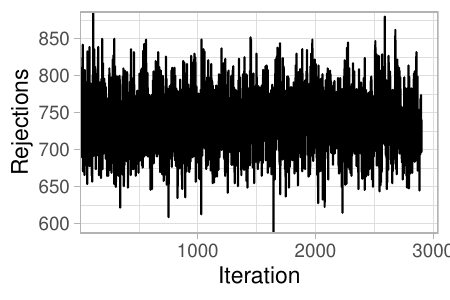}
\caption{$\epsilon_1 = 0.75$, $\epsilon_2 = 0.001$.}
\end{subfigure}
\begin{subfigure}[t]{0.32\textwidth}
\includegraphics[width=\textwidth]{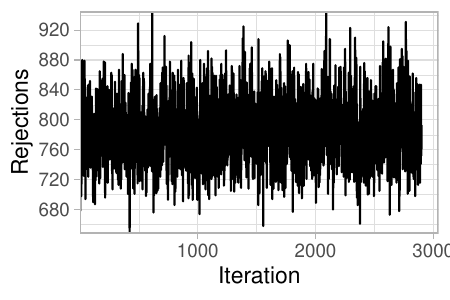}
\caption{$\epsilon_1 = 0.75$, $\epsilon_2 = 0.01$.}
\end{subfigure}

\begin{subfigure}[t]{0.32\textwidth}
\includegraphics[width=\textwidth]{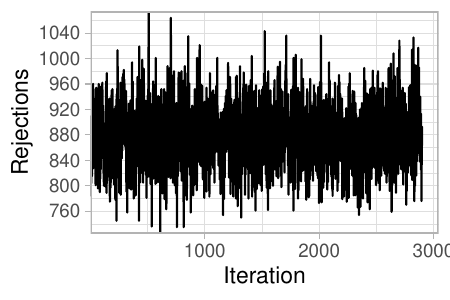}
\caption{$\epsilon_1 = 0.85$, $\epsilon_2 = 0.001$.}
\end{subfigure}
\begin{subfigure}[t]{0.32\textwidth}
\includegraphics[width=\textwidth]{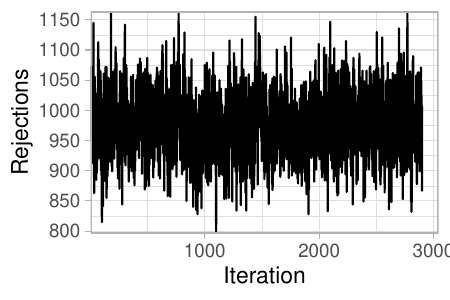}
\caption{$\epsilon_1 = 0.85$, $\epsilon_2 = 0.01$.}
\end{subfigure}
\end{figure}

Similar to the setup in Section~\ref{sec:saipe-analysis}, we compare the use of IMH, AMH and ARMS for the non-conjugate $\vec{\mu}$ draws with VWS under three settings: VWS with no self-tuning (VWS0), VWS with self-tuning (VWS1), and VWS with self-tuning for only the first 100 iterations (VWS2). Gibbs samplers with IMH and AMH are run for 30,000 iterations with a burn-in period of 28,000; all other variations use 3,000 iterations with a burn-in period of 1,000. Table~\ref{tab:unmatch-mixing} summarizes sampler performance including the number of rejections, total elapsed time in seconds, and quantiles of ESS and ESS per second for the elements of $\vec{\mu}$. Figure~\ref{fig:unmatch-ess} plots the empirical CDF of the ESS for elements of $\vec{\mu}$. Here the results displayed for VWS are based on VWS2 with $\epsilon_1 = 0.85$ and $\epsilon_2 = 0.01$. Figure~\ref{fig:unmatch-tune-ext0} plots total rejections by iteration for VWS0. Figure~\ref{fig:unmatch-tune-ext1} displays the total number of tuning adjustments, proposal regions, and rejections by iteration for VWS1.
Figure~\ref{fig:unmatch-tune-ext2} plots total rejections by iteration for VWS2.

\begin{figure}
\centering
\caption{Selected saved chains for $\vec{\mu}$ for VWS1 (light red) with $\epsilon_1 = 0.85$ and $\epsilon_2 = 0.01$ and IMH (dark blue) within Gibbs. The top row displays the three chains with lowest ESS using IMH. The bottom row shows the three chains with lowest ESS under VWS.}
\label{fig:unmatch-trace}
\includegraphics[width=0.32\textwidth]{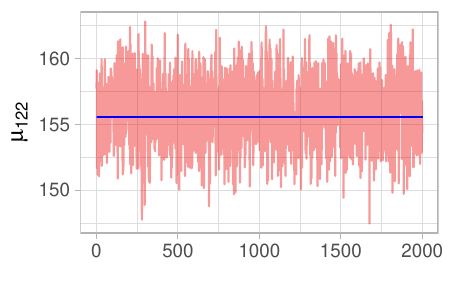}
\includegraphics[width=0.32\textwidth]{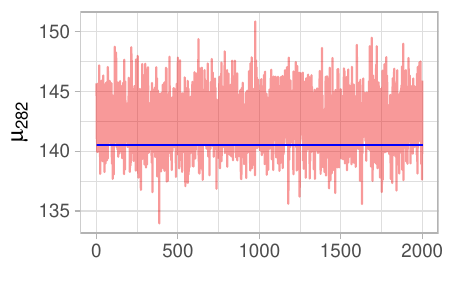}
\includegraphics[width=0.32\textwidth]{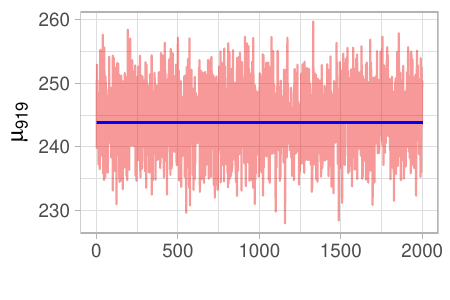}
\includegraphics[width=0.32\textwidth]{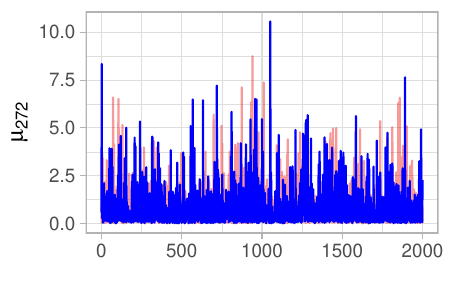}
\includegraphics[width=0.32\textwidth]{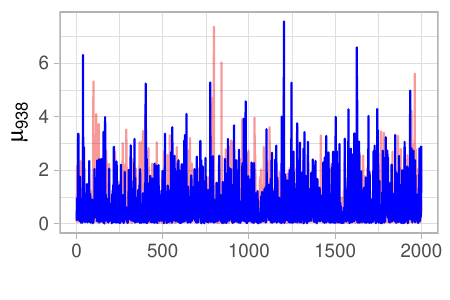}
\includegraphics[width=0.32\textwidth]{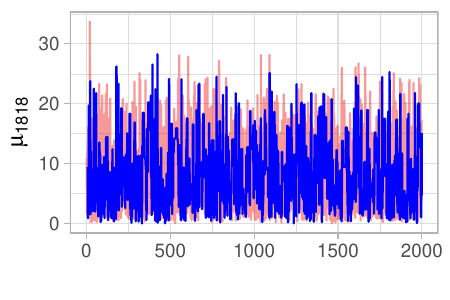}
\end{figure}

\begin{figure}
\centering
\caption{The three $\vec{\mu}$ chains with lowest ESS under AMH.}
\label{fig:unmatch-trace-amh}
\includegraphics[width=0.32\textwidth]{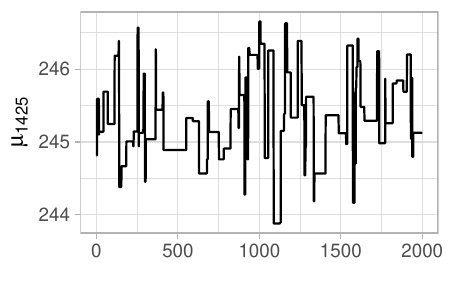}
\includegraphics[width=0.32\textwidth]{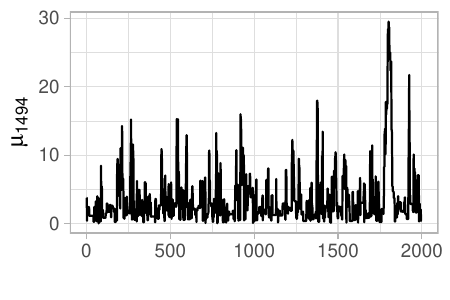}
\includegraphics[width=0.32\textwidth]{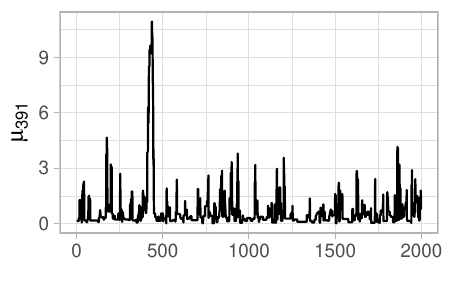}
\end{figure}

\begin{figure}
\centering
\caption{The three $\vec{\mu}$ chains with lowest ESS under ARMS.}
\label{fig:unmatch-trace-arms}
\includegraphics[width=0.32\textwidth]{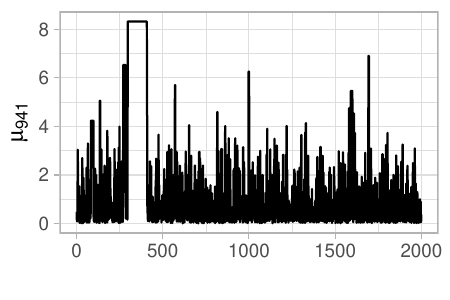}
\includegraphics[width=0.32\textwidth]{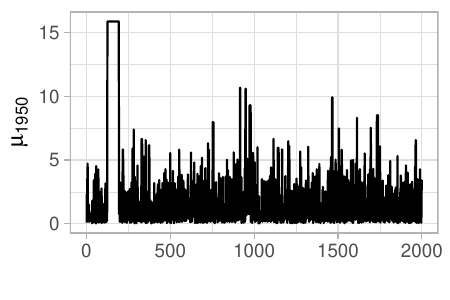}
\includegraphics[width=0.32\textwidth]{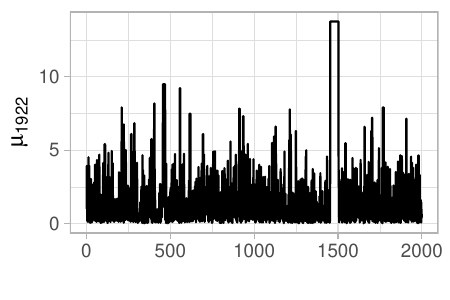}
\end{figure}

We see that IMH runs relatively very quickly, requiring only \red{7.68 seconds} in total; however, very low ESS is seen in some elements of $\vec{\mu}$ indicating poor mixing. In the worst cases, Figure~\ref{fig:unmatch-trace} shows that some elements fail to move at all. Mixing under AMH is 
an improvement over IMH but noticeably slow, especially in the worst cases seen in Figure~\ref{fig:unmatch-trace-amh}. In contrast to the joint SAE application, ARMS does not mix as well as VWS here; several periods are seen in Figure~\ref{fig:unmatch-trace-arms} where there is a lack of movement. Here, AMH outperforms ARMS in ESS per second because of the relatively low computational demand of AMH.

All three VWS approaches across all settings of $\epsilon_1$ and $\epsilon_2$ produce well-mixing chains with high ESS for all elements. Without tuning, VWS0 incurs the largest run time at \red{233.68} seconds. The use of $\epsilon_1 = 0.85$ and $\epsilon_2 = 0.01$ yields slightly better times than the other settings. Limiting tuning to the first 100 Gibbs iterations in VWS2 is seen to effectively reduce run time by about 60\% and increase ESS per second accordingly. Here the number of rejections increasing by only a relatively small amount, as seen in Table~\ref{tab:unmatch-summary} and in comparing Figures~\ref{fig:unmatch-tune-ext1} and \ref{fig:unmatch-tune-ext2}. The best VWS2 time of \red{60.24} seconds is on par with the ARMS run time of \red{54.18} seconds. Therefore, we consider self-tuning VWS---especially with a limited tuning period---to be an appealing method to sample from conditional \eqref{eqn:unmatch-target} within the proposed Gibbs sampler for this model.

\begin{table}
\centering
\caption{Comparison of samplers for $\vec{\mu}$ steps. The ``ESS'' column group displays three quantiles of effective sample size over the $m$ chains; ``ESS/sec'' divides these quantities by the elapsed time, which is given in seconds. Rejection counts are totaled over all Gibbs iterations (including burn-in period) and the $m$ chains.}
\label{tab:unmatch-mixing}
\begin{tabular}{lrlrrrrrrrr}
\toprule
\multicolumn{3}{c}{} &
\multicolumn{2}{c}{} & 
\multicolumn{3}{c}{ESS $\vec{\mu}$} &
\multicolumn{3}{c}{ESS/sec $\vec{\mu}$} \\
\cmidrule(lr){6-8}
\cmidrule(lr){9-11}
\multicolumn{1}{c}{} &
\multicolumn{1}{c}{$\epsilon_1$} &
\multicolumn{1}{c}{$\epsilon_2$} &
\multicolumn{1}{c}{Rejections} &
\multicolumn{1}{c}{Elapsed} &
\multicolumn{1}{c}{Min} &
\multicolumn{1}{c}{1\%} &
\multicolumn{1}{c}{2.5\%} &
\multicolumn{1}{c}{Min} &
\multicolumn{1}{c}{1\%} &
\multicolumn{1}{c}{2.5\%} \\
\cmidrule(lr){1-1}
\cmidrule(lr){2-3}
\cmidrule(lr){4-4}
\cmidrule(lr){5-5}
\cmidrule(lr){6-8}
\cmidrule(lr){9-11}
IMH  &  --- &   --- & 25,307,969 &   7.68 &     0 &     7 &    15 &  0.0 &  0.9 &  1.9 \\
AMH  &  --- &   --- & 38,550,163 &  13.82 &    45 &    79 &    96 &  3.2 &  5.7 &  6.9 \\
ARMS &  --- &   --- &     70,282 &  54.18 &    43 &   164 &   365 &  0.8 &  3.0 &  6.7 \\
VWS0 &  --- &   --- &  3,844,782 & 233.68 & 1,780 & 1,878 & 1,954 &  7.6 &  8.0 &  8.4 \\
\cmidrule(lr){1-1}
\cmidrule(lr){2-3}
\cmidrule(lr){4-5}
\cmidrule(lr){6-8}
\cmidrule(lr){9-11}
VWS1 & 0.75 & 0.001 &  2,129,083 & 104.50 & 1,744 & 1,877 & 1,953 & 16.7 & 18.0 & 18.7 \\
     & 0.75 & 0.01  &  2,271,331 & 107.49 & 1,723 & 1,885 & 1,931 & 16.0 & 17.5 & 18.0 \\
     & 0.85 & 0.001 &  2,668,191 & 106.63 & 1,722 & 1,862 & 1,916 & 16.2 & 17.5 & 18.0 \\
     & 0.85 & 0.01  &  2,817,569 & 101.86 & 1,778 & 1,855 & 1,924 & 17.5 & 18.2 & 18.9 \\
\cmidrule(lr){1-1}
\cmidrule(lr){2-3}
\cmidrule(lr){4-4}
\cmidrule(lr){5-5}
\cmidrule(lr){6-8}
\cmidrule(lr){9-11}
VWS2 & 0.75 & 0.001 &  2,223,101 &  65.35 & 1,710 & 1,887 & 2,000 & 26.2 & 28.9 & 30.6 \\
     & 0.75 & 0.01  &  2,367,489 &  61.90 & 1,721 & 1,902 & 2,000 & 27.8 & 30.7 & 32.3 \\
     & 0.85 & 0.001 &  2,644,831 &  62.70 & 1,779 & 1,875 & 1,959 & 28.4 & 29.9 & 31.2 \\
     & 0.85 & 0.01  &  2,937,942 &  60.24 & 1,639 & 1,861 & 1,912 & 27.2 & 30.9 & 31.7 \\
\bottomrule
\end{tabular}
\end{table}

\begin{table}
\centering
\caption{Posterior summary of saved draws for parameters in $\vec{\theta}$ using VWS1 with $\epsilon_1 = 0.85$ and $\epsilon_2 = 0.01$. The columns ``SD'', ``5\%'' and ``95\%'' represent the standard deviation, 5\% quantile, and 95\% quantile, respectively.}
\label{tab:unmatch-summary}
\begin{tabular}{rrrrr}
\toprule
\multicolumn{1}{c}{} &
\multicolumn{1}{c}{Mean} &
\multicolumn{1}{c}{SD} &
\multicolumn{1}{c}{5\%} &
\multicolumn{1}{c}{95\%} \\
\cmidrule(lr){1-1}
\cmidrule(lr){2-5}
$\beta_1$ &  1.0208 & 0.0520 &  0.9335 &  1.1041 \\
$\beta_2$ & -1.0276 & 0.0448 & -1.1041 & -0.9585 \\
$\tau^2$  &  1.4949 & 0.0899 &  1.3560 &  1.6472 \\
\bottomrule
\end{tabular}
\end{table}

\begin{table}
\centering
\caption{Effective sample size of parameters in $\vec{\theta}$ under varying samplers for the $\vec{\mu}$ step.}
\label{tab:unmatch-theta-ess}
\begin{tabular}{lrlrrr}
\toprule
\multicolumn{1}{c}{} &
\multicolumn{1}{c}{$\epsilon_1$} &
\multicolumn{1}{c}{$\epsilon_2$} &
\multicolumn{1}{c}{$\beta_1$} &
\multicolumn{1}{c}{$\beta_2$} &
\multicolumn{1}{c}{$\tau^2$} \\
\cmidrule(lr){1-1}
\cmidrule(lr){2-3}
\cmidrule(lr){4-6}
IMH  &  --- &   --- & 142 & 253 & 136 \\
AMH  &  --- &   --- & 168 & 257 & 240 \\
ARMS &  --- &   --- & 168 & 257 & 240 \\
VWS0 &  --- &   --- & 150 & 271 & 188 \\
\cmidrule(lr){1-1}
\cmidrule(lr){2-3}
\cmidrule(lr){4-6}
VWS1 & 0.75 & 0.001 & 203 & 252 & 239 \\
     & 0.75 & 0.01  & 189 & 330 & 215 \\
     & 0.85 & 0.001 & 194 & 477 & 166 \\
     & 0.85 & 0.01  & 183 & 264 & 222 \\
\cmidrule(lr){1-1}
\cmidrule(lr){2-3}
\cmidrule(lr){4-6}
VWS2 & 0.75 & 0.001 & 216 & 304 & 220 \\
     & 0.75 & 0.01  & 222 & 348 & 191 \\
     & 0.85 & 0.001 & 184 & 329 & 278 \\
     & 0.85 & 0.01  & 209 & 392 & 208 \\
\bottomrule
\end{tabular}
\end{table}

\begin{figure}
\centering
\caption{Credible intervals of elements of $\vec{\mu}$ whose coverage of the true $\vec{\mu}$ changed from use of IMH versus VWS. Coordinates correspond to changes in the lower and upper endpoints of the interval. Larger shifts are labeled with the ESS value from IMH. Not shown in (\subref{fig:unmatch-to-uncovg}) is a point at $(17.25, 40.46)$ with ESS 339.}
\label{fig:unmatch-covg}
\begin{subfigure}[t]{0.80\textwidth}
\includegraphics[width=\textwidth]{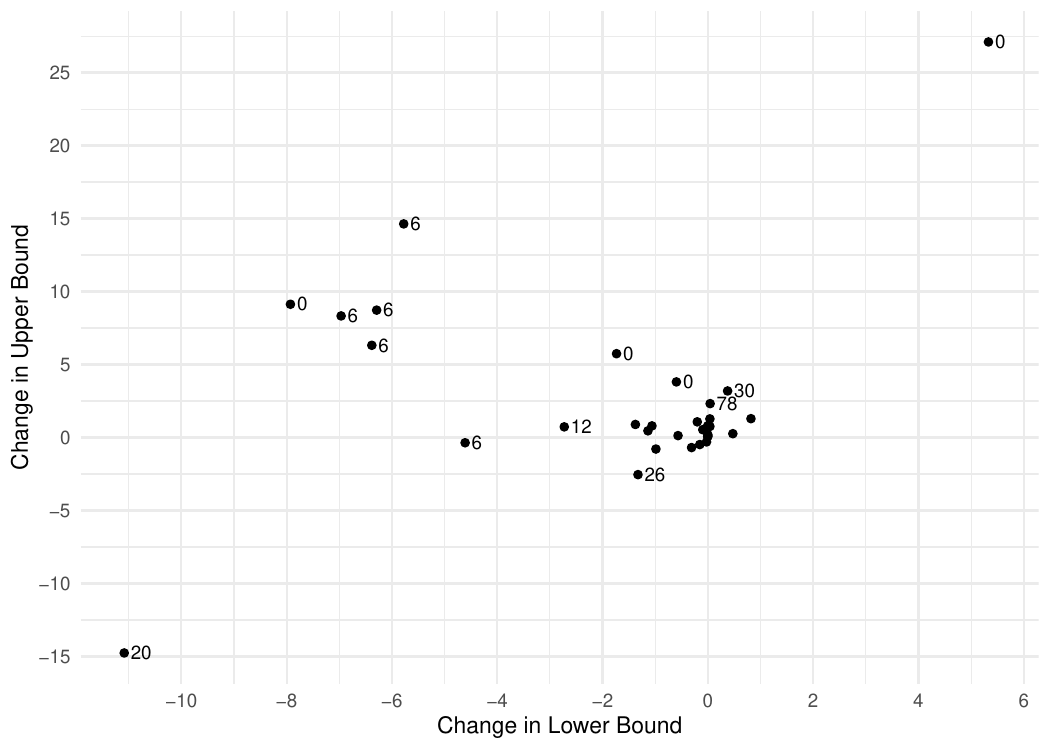}
\caption{Elements uncovered under IMH but covered under VWS.}
\label{fig:unmatch-to-covg}
\end{subfigure}
\begin{subfigure}[t]{0.80\textwidth}
\includegraphics[width=\textwidth]{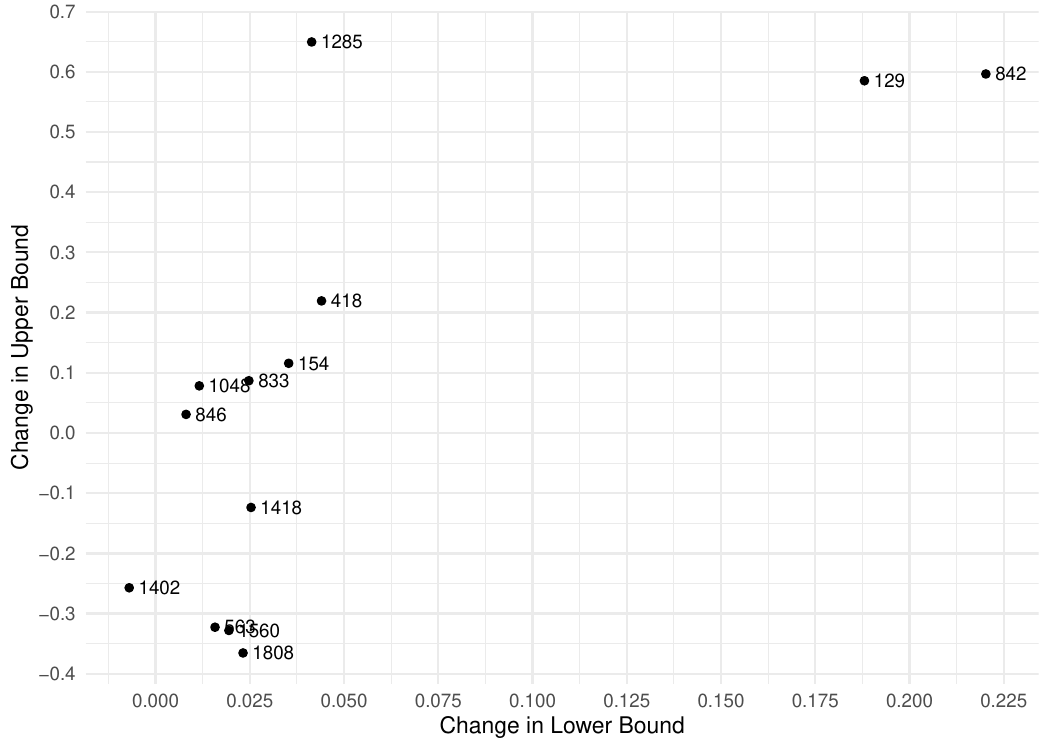}
\caption{Elements covered under IMH but uncovered under VWS.}
\label{fig:unmatch-to-uncovg}
\end{subfigure}
\end{figure}

Figure~\ref{fig:unmatch-covg} considers credible intervals for elements of $\vec{\mu}$ with 90\% nominal coverage, and their empirical coverage of the true latent values. Considering intervals computed under both IMH and VWS, Figure~\ref{fig:unmatch-to-covg} displays the \red{32} observations where the true value of $\mu_i$ is uncovered under IMH and covered under VWS. The horizontal and vertical coordinates of each point correspond to the change in lower and upper endpoint of the interval, respectively, from IMH to VWS. Larger shifts are labeled with the ESS value from IMH. Similarly, Figure~\ref{fig:unmatch-to-uncovg} displays the \red{14} observations where $\mu_i$ is covered under IMH and uncovered under VWS. We see that many of the observations in Figure~\ref{fig:unmatch-to-covg} were also ones which mixed the most poorly in IMH. On the other hand, for the observations Figure~\ref{fig:unmatch-to-uncovg}, the change in coverage appears to be from inherent randomness in the MCMC and not poor mixing.

\clearpage


\end{document}